\documentclass[12pt]{article}
\usepackage{amsmath,amssymb,amsfonts}
\usepackage[noblocks]{authblk}
\usepackage{booktabs}
\usepackage{multirow}
\usepackage{abstract}
\usepackage{booktabs}
\usepackage{authblk}
\usepackage{indentfirst}
\usepackage{cases} 
\usepackage{diagbox,multirow,makecell}
\usepackage{multicol}
\usepackage{graphicx}
\usepackage{threeparttable}
\usepackage{natbib}
\usepackage{apalike} 
\usepackage{pgf}
\usepackage{tikz}
\usepackage{mathtools}
\usetikzlibrary{arrows, decorations.pathmorphing, backgrounds, positioning, fit, petri, automata,shapes}
\usepackage[colorlinks,linkcolor=black,
anchorcolor=black,
citecolor=black,
 urlcolor=blue]{hyperref}
\usepackage{url}
\newtheorem{assumption}{Assumption}
\newtheorem{thm}{Theorem}

\begin{document}
	\title{An IRT-based Model for Omitted and Not-reached Items}
     \author{Jinxin Guo}
     \author{Xin Xu}
     \affil{Northeast Normal University}
	\date{}
	\maketitle

	\setlength{\baselineskip}{2em}
	
\begin{abstract}
Missingness is a common occurrence in educational assessment and psychological measurement. It could not be casually ignored as it may threaten the validity of the test if not handled properly. Considering the difference between omitted and not-reached items, we developed an IRT-based model to handle these missingness. In the proposed method, not-reached responses are captured by the cumulative missingness. Moreover, the nonignorability is attributed to the correlation between ability and person missing trait. We proved that its item parameters estimate under maximum marginal
likelihood (MML) estimation is consistent. We further proposed a Bayesian estimation procedure using MCMC methods to estimate all the parameters. The simulation results indicate that the model parameters under the proposed method are better recovered than that under listwise deletion, and the nonignorable model fits the simulated nonignorable nonresponses better than ignorable model in terms of Bayesian model selection. Furthermore, the Program for International Student Assessment (PISA) data set was analyzed to further illustrate the usage of the proposed method.\\	

\noindent\textbf{Keywords}: missing data mechanism, omitted, not-reached, nonignorable, ignorable, PISA.
\end{abstract}
\section{Introduction}
Missing data is always unavoidable in many studies, including educational assessment and psychological measurement \citep{rose2017modeling,yuan2018missing}. Recently modeling missing data mechanism has gained increasing prominence and been widely considered in order to get a more reliable evaluation. Actually, missingness would occur under many conditions. For example, test takers may fail to reach some items due to time limits. Or sometimes they may tend to omit some items for individual reasons, such as their abilities and item preference. If these missing responses could not be dealt with properly, it would bring biased parameter estimation and further threaten the validity of tests \citep{pohl2014dealing,rose2015commonalities}.
	
To better tackle the problem of missing data, \citet{rubin1976inference} and \citet{little2014statistical} have defined three kinds of missingness: \textquotedblleft missing completely at random\textquotedblright (MCAR), \textquotedblleft missing at random\textquotedblright (MAR), and \textquotedblleft not missing at random\textquotedblright (NMAR). 

Suppose that there are $N$ examinees and $J$ items in the test. Let $Y_{ij}$ denote the dichotomous response variable of examinee $i\in\{1,\cdots,N\}$ to item $j\in\{1,\cdots,J\}$, where $Y_{ij}=1$ means that examinee $ i $ answer item $j$ correctly while $Y_{ij}=0$ otherwise. Let
$Y_{ij}=(Y_{ij}^{\text{obs}},Y_{ij}^{\text{mis}})$, 
where $Y_{ij}^{\text{obs}} $ is observed and $Y_{ij}^{\text{mis}}  $ is missing. And $\boldsymbol{Y}=\{Y_{ij}\}_{N\times J}$ is the complete response matrix which can be decomposed into an observed part $\boldsymbol{Y^{\text{obs}} }$ and a missing part $\boldsymbol{Y^{\text{mis}} }$. Denote by $\boldsymbol{R}=\{R_{ij}\}_{N\times J}$ the missing indicator matrix, where the missing variable $R_{ij}$ can be defined as
\[
R_{ij}=\left\{
\begin{tabular}
[c]{ll}%
0, & if $Y_{ij}$ is observed,\\
& \\
1, & if $Y_{ij}$ is missing.
\end{tabular}
\right.
\]

Actually, the three categories of missingness can be characterized by the conditional distribution of $ \boldsymbol{R} $ given $ \boldsymbol{Y} $, denoted by $ P(\boldsymbol{R}|\boldsymbol{Y}, \boldsymbol{\Omega}) $, where $ \boldsymbol{\Omega} $ is the unknown parameters set.  The missingness is MCAR if the distribution of $\boldsymbol{R}$ does not depend on the response data $ \boldsymbol{Y} $, either observed or unobserved, which can be formulated as
\[ P(\boldsymbol{R}|\boldsymbol{Y}, \boldsymbol{\Omega})= P(\boldsymbol{R}| \boldsymbol{\Omega}) \text{ for all } \boldsymbol{Y}, \boldsymbol{\Omega}.\]
MAR is the situation in which missingness is independent of missing response given the observed ones, which can be written as \[ P(\boldsymbol{R}|\boldsymbol{Y}, \boldsymbol{\Omega})= P(\boldsymbol{R}| \boldsymbol{Y^{\text{obs}} },\boldsymbol{\Omega})\text{ for all } \boldsymbol{Y^{\text{mis}} }, \boldsymbol{\Omega}.\] The missingness of MCAR or MAR is also called ignorable or uninformative \citep{schafer2002missing,rose2015commonalities}. However, NMAR is obviously distinct from the above two kinds of missingness. It is not independent of missing response given the observed responses, which is also called nonignorable or informative. For example, in the context of IRT, test takers may fail to answer some items because their abilities are too low to answer correctly. This kind of missing responses can be viewed as NMAR.
	
In fact, there exist several methods to handle missing item responses. One of the most direct and simplest approaches is listwise deletion which is also the default method dealing with missing data in some statistical softwares, such as SPSS and SAS. In this method, all cases with missing responses would be deleted. The method is very direct and effective when the missing data response rate is small. However, if the proportion of missing data is high, especially when the missingness is nonignorable, this method would cause bias and thus lead to errors in statistical inference \citep{rose2015commonalities,rose2017modeling,wu2017bayesian}.	

So we paid more attention to modeling missing responses for nonignorable missingness. The general methods are typically based on the joint distribution  of $\boldsymbol{R}$ and $\boldsymbol{Y}$ was constructed. Two commonly used joint models are selection models (SLM; \citealp{heckman1976common}) and pattern mixture models (PMM; \citealp{little1993pattern}).
SLM is based on the following factorization:
\begin{equation}
\label{slm}
P(\boldsymbol{R},\boldsymbol{Y}| \boldsymbol{\Omega})=P(\boldsymbol{R}|\boldsymbol{Y},\boldsymbol{\Omega})P(\boldsymbol{Y}| \boldsymbol{\Omega})
\end{equation}
And PMM can be written as:
\begin{equation}
P(\boldsymbol{R},\boldsymbol{Y}| \boldsymbol{\Omega})=P(\boldsymbol{Y}|\boldsymbol{R},\boldsymbol{\Omega})P(\boldsymbol{R}| \boldsymbol{\Omega})
\end{equation}

 Based on these joint models, several methods have been proposed by researchers. For example, \citet{holman2005modelling} and \citet{glas2015nonignorable} introduced an IRT model for omitted items based on PMM that could simultaneously estimate IRT item parameters and the parameter about the propensity of missing data. \citet{rose2010modeling} derived multidimensional IRT (MIRT) to handle nonignorable item nonresponses, which was believed to have originated from general SLM \citep{rose2013item}. However, these methods could only be applied to omitted responses. To differentiate the omitted and not-reached items, latent regression models (LRMs) were proposed to model omitted and not-reached items \citep{rose2017modeling}.

Specifically, not-reached items (also called \textquotedblleft dropout\textquotedblright) occurs when the test takers fail to reach some items at the end of test and omitted items (also called \textquotedblleft intermittent\textquotedblright) refer to the situation where they skip one or more items and then answer the next one.

Motivated by the previous methods, this paper proposed an approach to model the omitted and not-reached items on the basis of SLM. In details, the effects of previous nonresponses on current item are modeled for not-reached items. And the correlation between ability parameter and latent person missing parameter is employed to clarify whether the missingness is nonignorable or ignorable.
	
The remainder of this paper was organized as follows. In Section 2, we presented the proposed method to handle binary missing item responses. The MML estimation of the item parameters and related consistency results were given in Section 3, followed by the Bayesian estimation using MCMC method. To evaluate parameter recovery and model selection, two simulation studies were conducted in Section 4. In Section 5, we carried out a detailed analysis of PISA data set to illustrate the usage of the proposed method. Finally, some issues that need to be resolved were addressed and further research directions were discussed in Section 6.

\section{Handling omitted and not-reached items with IRT-based model }
	\subsection{Two parameter IRT models}
In the proposed method, two-parameter IRT models
	 were employed to model the item response data. The probability of examinee $i$ correctly answering item $j$ can be expressed as
	\begin{equation}
	\label{2pno}
	p_{ij}=P(Y_{ij}=1\left\vert \theta_{i},a_{j},b_{j}\right.  )=F\left(
	a_{j}\left(  \theta_{i}-b_{j}\right)  \right), 
	\end{equation}
	where $\theta_{i}$ denotes the ability parameter for the $i$th individual; $a_{j}$ and $b_{j}$ are the discrimination and difficulty parameters of item $j$, respectively. And $F\left(\cdot\right) $ is a cumulative distribution function (CDF) of standard normal or standard logistic distribution. In details, probit link could yield two-parameter  
	normal ogive (2PNO) IRT model
	, that is, $ p_{ij}=\Phi\left(
	a_{j}\left(  \theta_{i}-b_{j}\right)  \right) $, where $\Phi(  \cdot )$ is the CDF of standard normal distribution. Similarly, when $F\left(\cdot\right) $ is the CDF of standard logistic distribution, that is, logit link is employed
	, it follows that $ \text{logit}(p_{ij})=	a_{j}\left(  \theta_{i}-b_{j}\right)  $. Therefore, $ p_{ij}=1/\{1+\exp\left(
		-a_{j}\left(  \theta_{i}-b_{j}\right)  \right)\} $, which is called two-parameter  
	logistic (2PL) IRT model. 
	
\subsection{Modeling missing data mechanism}
 Considering the difference between omitted and not-reached items, different effect indexes were applied in the proposed method. Especially, not-reached responses are characterized by the cumulative missingness. Motivated by the idea of IRT models, missingness is captured by latent missing trait from two perspectives: item and person. On the basis of this, the missing data process can be modeled as 
\begin{equation}\label{pir}
\begin{split}
	\pi_{ij}=&P\left(  R_{ij}=1\left\vert \tau_{i},\zeta_{j},\boldsymbol{R}_{i,j-1},\boldsymbol{Y}_{ij},\boldsymbol{\gamma} \right.  \right) \\
	 = &G\left(  \gamma_{0}-\tau
	_{i}+\zeta_{j}+g\left(  \boldsymbol{R}_{i,j-1},\gamma_{1}\right)  +l\left(
	\boldsymbol{Y}_{ij},\gamma_{2}\right)  \right)  ,
\end{split}
\end{equation}
where $G\left(\cdot\right) $ is similar to $F\left(\cdot\right) $ in Equation \eqref{2pno}, specified by probit or logit link, $\gamma_{0}$ is the intercept parameter with constraint of $ \gamma_{0}<0$ that could influence the baseline probability of missingness, $\tau_{i}$ is the person missing parameter that measures the individual latent trait to nonresponses, and $\zeta_{j}$ is the item missing parameter that represents the inclination of missingness caused by item. Moreover, for the $i$th examinee, $\boldsymbol{R}_{i,j-1}=\left(  R_{i1},R_{i2},...,R_{i,j-1}\right)$ is the previous missing vector before $j$th item and $\boldsymbol{Y}_{ij}=\left(  Y_{i1},Y_{i2},...,Y_{ij}\right)$ is responses vector of the first $j$ items. Moreover, $l\left(\cdot\right)$ is a
function of the responses vector that characterizes the effect from response variables. And we simply took $l\left(  \boldsymbol{Y}_{ij},\gamma_{2}\right)  =\gamma_{2}Y_{ij}$, where $ \gamma_{2}<0 $. In addition, $g\left( \cdot\right)$ is a function of the missing indicators, which denotes the impact of previous nonresponses on current item. We set $g\left( \cdot\right) =0$ when
$j=1$ since there are no previous missing responses. Specially, in the proposed method, $g\left(  \boldsymbol{R}_{i,j-1}
,\gamma_{1}\right)=\gamma_{1}\sum\limits_{h=2}^{j-1}{R}_{i,h}$ with $\gamma_{1}>0$ was chosen to model the effect of previous missingness on the current item. Actually, the statistic $\sum\limits_{h=2}^{j-1}{R}_{i,h}$ exactly captures the missingness of not-reached items, and reduces the number of nuisance parameters for modeling the missing data mechanism.  

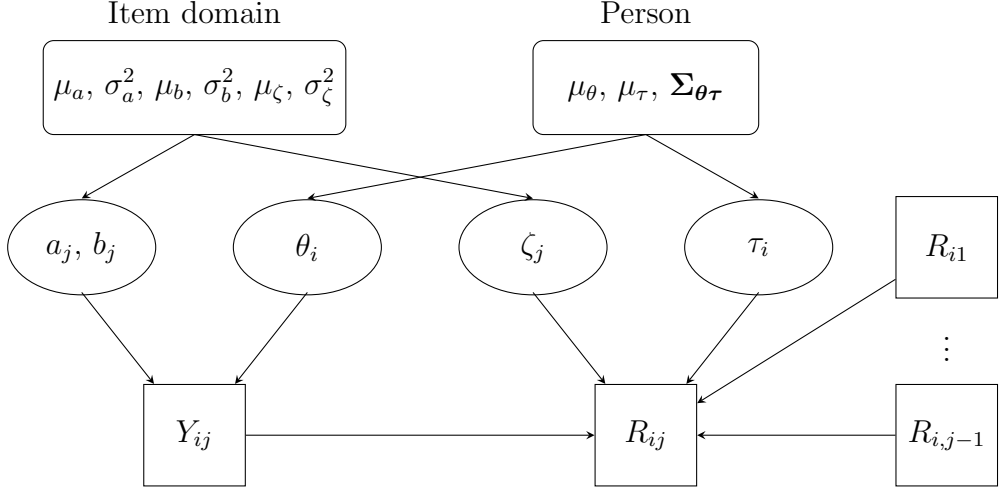
\begin{figure}[htb]
			\tikzstyle{node1} = [rectangle,minimum width =1.35cm, minimum height = 1.35cm,text centered, draw = black]
			\tikzstyle{node} = [rectangle,minimum width =3cm, minimum height = 1.25cm,rounded corners,text centered, draw = black]
			\tikzstyle{arrow} = [thin, ->, >= stealth]
			\begin{tikzpicture}[node distance = 1.5cm]
			\node (itemdomain1)at (-7.5,3.125)    {Item domain};
			\node[node] (itemdomain)at (-7.5, 2.125)    [draw]{$ \mu_{a} $, $ \sigma_{a}^{2} $, $ \mu_{b} $, $ \sigma_{b}^{2} $, $ \mu_{\zeta} $, $ \sigma_{\zeta}^{2} $ };
				\node (person1)at (-1.5, 3.125)    {Person};
			\node[node](person)at (-1.5,2.125)    [draw]{$ \mu_{\theta} $, $ \mu_{\tau} $, $\boldsymbol{\Sigma_{\theta \tau}}$ };

			\node (itempar)at (-9, 0)    {$ a_{j} $, $ b_{j}$};
			\draw (-9,0) ellipse (28pt and 18pt);
			\node (ability)at (-6, 0)    {$ \theta_{i} $};
			\draw (-6,0) ellipse (28pt and 18pt);
			\node(itemmissing)at (-3, 0)    {$ \zeta_{j} $};
			\draw (-3,0) ellipse (28pt and 18pt);
			\node (personmissing)at (0,0)    {$ \tau_{i} $};
			\draw (0,0) ellipse (28pt and 18pt);
			\node[node1] (response)at (-7.5, -2.5)    [draw]{
				$ Y_{ij} $};
			\node[node1] (missing)at (-1.5, -2.5)    [draw]{
       	$ R_{ij} $};
       	\node[node1] (r1)at (2.5, 0)    [draw]{
       	$ R_{i1} $};
      	\node  (vdots)at (2.5, -1.25)   {
       	$ \vdots $};
       \node[node1] (rj-1)at (2.5, -2.5)    [draw]{
       	$ R_{i,j-1} $};
			\draw[arrow] (-7.5,1.5) to (-9,0.65);
			\draw[arrow] (-7.5,1.5) to (-3, 0.65);
			\draw[arrow] (-1.5,1.5) to (-6,0.65);
			\draw[arrow] (-1.5,1.5) to (0,0.65);
			\draw[arrow] (-9, -0.6) to (response);   
			\draw[arrow] (-6, -0.6) to (response);
			\draw[arrow] (-3, -0.6) to (missing); 
			\draw[arrow] (0, -0.6) to (missing); 
			\draw[arrow] (response) to (missing) ;
				\draw[arrow] (r1) to (missing) ;
			\draw[arrow] (rj-1) to (missing) ;		
			\end{tikzpicture}
	
		\caption{The graphical representation of the proposed method.\label{graph}}
	\end{figure}

In the proposed method, the ability parameter $\theta$ in IRT model and person missing parameter $\tau$ in missing part were assumed to be bivariate joint normally distributed with mean vector $\boldsymbol{\mu}=\left(
\begin{tabular}
[c]{l}
$\mu_{\theta}$\\
$\mu_{\tau}$
\end{tabular}
\ \right) $ and covariance matrix $\boldsymbol{\Sigma}=\left(
\begin{tabular}{cc}
$\sigma^{2}_{\theta}$&$\sigma_{\theta \tau}$  \\ 
$\sigma_{\theta \tau}$& $\sigma^{2}_{\tau}$ \\ 
\end{tabular} 
\ \right) $. A graphical representation of the proposed method is present in Figure~\ref{graph}.

 To guarantee the model identification, $\mu_{\theta}$ and $\mu_{\tau}$ are set to 0, and $\sigma^{2}_{\theta}$ was fixed to 1 \citep{browne2006mcmc}.
Note that the generation of the nonignorable mechanism is attributed to the
correlation between  $\theta$ and $\tau$. To be more specific, if the missingness is ignorable, $ \tau$ is independent of $\theta$, therefore, $p\left(
\theta,\tau\left\vert \boldsymbol{\mu},\boldsymbol{\Sigma}\right. \right)=p(\theta|\mu_{\theta},\sigma^{2}_{\theta})p(\tau|\mu_{\tau},\sigma^{2}_{\tau})$. If the missing data depends on both the latent ability $\theta$ and the person missing parameter $\tau$, the missingness is nonignorable. So the proposed method could handle both ignorable and nonignorable missingness. 	

\subsection{The likelihood function}
	Let $\boldsymbol{\theta}=(\theta_{1},\cdots,\theta_{N})$, $\boldsymbol{\tau}=(\tau_{1},\cdots,\tau_{N})$, $\boldsymbol{a}=(a_{1},\cdots,a_{J})$, $\boldsymbol{b}=(b_{1},\cdots,b_{J})$, $\boldsymbol{\zeta}=(\zeta_{1},\cdots,\zeta_{J})$, and $\boldsymbol{\gamma}=(\gamma_{0},\gamma_{1},\gamma_{2})$ . And the parameter set can be written as $\boldsymbol{\Omega} = (\boldsymbol{\theta},\boldsymbol{\tau},\boldsymbol{a},\boldsymbol{b},\boldsymbol{\zeta},\boldsymbol{\gamma},\sigma_{\theta \tau}, \sigma^{2}_{\tau})$. Denote by $\boldsymbol{y}_{ij}=(y_{i1},\cdots,y_{ij})$ and $\boldsymbol{r}_{ij}=(r_{i1},\cdots,r_{ij})$ the observation vectors of 
 $\boldsymbol{Y}_{ij}=(Y_{i1},\cdots,Y_{ij})$ 
 and $\boldsymbol{R}_{ij}=(R_{i1},\cdots,R_{ij})$.
	
	A sequence of one-dimensional conditional distributions modeling method proposed by \citet{ibrahim1999missing} was employed to construct the conditional joint distribution of $\boldsymbol{R_{i\cdot}}$ given $ \boldsymbol{Y_{iJ}} $, which can be written as
	\begin{equation}
	\begin{split}
	P\left(  \boldsymbol{R_{iJ}}=\boldsymbol{r_{iJ}}\left\vert
	\boldsymbol{\gamma},\tau_{i},\boldsymbol{\zeta
	},\boldsymbol{Y}_{iJ}\right.  \right)  =&\prod\limits_{j=1}^{J}P\left(  R_{ij}=r_{ij}\left\vert \tau
	_{i},\zeta_{j},\boldsymbol{Y_{ij}}\right.,\boldsymbol{R}_{i,j-1}  \right)\\
	=&\prod\limits_{j=1}^{J}%
	\pi_{ij}^{\mathrm{I}(  r_{ij}=1)  }\left(  1-\pi_{ij}\right)  ^{\mathrm{I}(  r_{ij}=0)  },
	\end{split}
	\end{equation}
	where $ \mathrm{I}(\cdot) $ is the indicator function and $\pi_{ij}$ is given by Equation \eqref{pir}.
	
    Based on Equation \eqref{slm}, the likelihood function of complete data could be given by 
\begin{equation}
	\label{likelihood}
	\begin{split}
	L(\boldsymbol{\Omega}|\boldsymbol{y},\boldsymbol{r})=&\prod\limits_{i=1}^{N}P(  \boldsymbol{Y}_{iJ}=\boldsymbol{y}_{iJ}|
	\boldsymbol{a},\boldsymbol{b},\theta_{i})P(  \boldsymbol{R}_{iJ}=\boldsymbol{r}_{iJ}|
	\boldsymbol{\gamma},\tau_{i},\boldsymbol{\zeta
	},\boldsymbol{Y}_{iJ})\\
	=&\prod\limits_{i=1}^{N}\prod\limits_{j=1}^{J}P\left(  Y_{ij}=y_{ij}\left\vert
	\theta_{i},a_{j},b_{j}\right.  \right)  P\left(  R_{ij}=r_{ij}\left\vert \tau
	_{i},\zeta_{j},\boldsymbol{Y}_{ij}\right.,\boldsymbol{R}_{i,j-1}  \right) \\
	=&\prod\limits_{i=1}^{N}\prod\limits_{j=1}^{J}p_{ij}^{y_{ij}}(1-p_{ij})^{1-y_{ij}}\pi_{ij}^{r_{ij}}(1-\pi_{ij})^{1-r_{ij}}.
	\end{split}\end{equation}
where $p_{ij} $ is given by Equation \eqref{2pno}. Note that if $Y_{ij}$ in Equation \eqref{likelihood} is missing, it would be imputed from Bernoulli$\left(  q_{ij}\right)$, and $q_{ij}$ can be computed as
\begin{equation}\label{qij}
\begin{split}
q_{ij}=P(Y_{ij}^{\text{mis}}=1|a_{j},b_{j},\zeta_{j},\theta_{i},\tau_{i},\boldsymbol{\gamma})
=\frac{p_{ij}\pi_{ij}^{11}}{p_{ij}\pi_{ij}^{11}+(  1-p_{ij})\pi_{ij}^{10} },
\end{split}
\end{equation}
where 
\begin{equation}
\begin{split}
 \pi_{ij}^{11}=&P(  R_{ij}=1|\tau
_{i},\zeta_{j},\boldsymbol{\gamma},\boldsymbol{R}_{i,j-1},Y_{ij}=1),\\
\pi_{ij}^{10}=&P(R_{ij}=1|\tau
_{i},\zeta_{j},\boldsymbol{\gamma},\boldsymbol{R}_{i,j-1},Y_{ij}=0). 
\end{split}
\end{equation}

Integrating over all
the missing item responses $\boldsymbol{y^{\text{mis}} }$ finally yields 
\begin{equation}
\label{obslikelihood}
\begin{split}
L(\boldsymbol{\Omega}|\boldsymbol{y^{\text{obs}}},\boldsymbol{r})
&=\sum_{\boldsymbol{y^{\text{mis}} }} \prod\limits_{i=1}^{N}\prod\limits_{j=1}^{J}
P\left(  Y_{ij}=(y_{ij}^{\text{obs}},y_{ij}^{\text{mis}})\left\vert
\theta_{i},a_{j},b_{j}\right.  \right)\\
&\cdot P\left(  R_{ij}=r_{ij}\left\vert \tau
_{i},\zeta_{j},\boldsymbol{Y}_{ij}\right.,\boldsymbol{R}_{i,j-1}  \right)P(Y_{ij}=y_{ij}^{\text{mis}}|a_{j},b_{j},\zeta_{j},\theta_{i},\tau_{i},\boldsymbol{\gamma})\\
&=\sum_{\boldsymbol{y^{\text{mis}} }}\prod\limits_{i=1}^{N}\prod\limits_{j=1}^{J}p_{ij}^{y_{ij}^{\text{obs}}}(1-p_{ij})^{1-y_{ij}^{\text{obs}}}\pi_{ij}^{r_{ij}}(1-\pi_{ij})^{1-r_{ij}}q_{ij}^{y_{ij}^{\text{mis}}}(1-q_{ij})^{1-y_{ij}^\text{mis}},
\end{split}\end{equation}

where $ y_{ij}^{\text{obs}} $ is the observation of $ Y_{ij}^{\text{obs}} $ and $ y_{ij}^{\text{mis}} $ is the imputation of $ Y_{ij}^{\text{mis}} $.

\section{Estimation of model parameters and consistency results}

 \subsection{MML estimation and consistency results }
 We first presented the MML estimation of IRT item parameters and item missing parameter. By integrating over the ability and person missing parameters, the marginal likelihood function $L(\boldsymbol{a},\boldsymbol{b},\boldsymbol{\zeta}|\boldsymbol{y^{\text{obs}}},\boldsymbol{r})  $ could by given by 
 \begin{equation*}
 \sum_{\boldsymbol{y^{\text{mis}} }}\prod\limits_{i=1}^{N}\prod\limits_{j=1}^{J}\iint p_{ij}^{y_{ij}^{\text{obs}}}(1-p_{ij})^{1-y_{ij}^{\text{obs}}}\pi_{ij}^{r_{ij}}(1-\pi_{ij})^{1-r_{ij}}q_{ij}^{y_{ij}^{\text{mis}}}(1-q_{ij})^{1-y_{ij}^\text{mis}}p(
  \theta,\tau|\boldsymbol{\mu},\boldsymbol{\Sigma})\,d\theta\,d\tau,
 \end{equation*}
where $\sigma_{\theta \tau}$ in $\boldsymbol{\Sigma}$ is assumed to be known. The MML estimation of $ (\boldsymbol{a},\boldsymbol{b},\boldsymbol{\zeta}) $ should satisfy:
 \[(\hat{\boldsymbol{a}},\hat{\boldsymbol{b}},\hat{\boldsymbol{\zeta}})=\mathop {\arg \max }\limits_{\boldsymbol{a},\boldsymbol{b},\boldsymbol{\zeta}}l(\boldsymbol{a},\boldsymbol{b},\boldsymbol{\zeta}|\boldsymbol{y^{\text{obs}}},\boldsymbol{r})\]
 where $ l(\boldsymbol{a},\boldsymbol{b},\boldsymbol{\zeta}|\boldsymbol{y^{\text{obs}}},\boldsymbol{r})=\log L(\boldsymbol{a},\boldsymbol{b},\boldsymbol{\zeta}|\boldsymbol{y^{\text{obs}}},\boldsymbol{r})$ is the log likelihood.
 In practice, the log likelihood equations for the item parameters could be derived 
 using the Newton–Raphson algorithm, EM algorithm, or a combination of the two \citep{bock1981marginal}. Therefore, the MML estimation of $ (\boldsymbol{a},\boldsymbol{b},\boldsymbol{\zeta}) $ could be easily obtained. Actually the MML estimation is consistent under some assumptions.

\begin{assumption}\label{c1}
	For sufficiently small $A_{1}>0$ and for sufficiently large $ A_{2}>0$, $ B>0 $, $ C>0 $, the following integrals are finite:
	\begin{equation*}
	\begin{split}
&	m_{1}(A_{1},B) =-\iint  \log F(A_{1}(\theta-B)) p(
\theta,\tau|\boldsymbol{\mu},\boldsymbol{\Sigma})\,d\theta\,d\tau 
,	\\
&	m_{2}(A_{2},B) =-\iint \log [1-F(A_{2}(\theta+B))]p(
\theta,\tau|\boldsymbol{\mu},\boldsymbol{\Sigma})\,d\theta\,d\tau
,	\\&	m_{3}(C)=-\iint\log G(c-\tau-C)p(
\theta,\tau|\boldsymbol{\mu},\boldsymbol{\Sigma})\,d\theta\,d\tau
,	\\&	m_{4}(C) =-\iint\log [1-G(c-\tau-C)]p(
\theta,\tau|\boldsymbol{\mu},\boldsymbol{\Sigma})\,d\theta\,d\tau
,	\end{split}
	\end{equation*}
	where $ c $ is a known finite constant.
\end{assumption}
\begin{assumption}\label{c2}
Given known $\boldsymbol{\gamma}$, $ (\boldsymbol{a},\boldsymbol{b},\boldsymbol{\zeta}) $ are identifiable.  
\end{assumption}

 \begin{thm}\label{thm1}
The MML estimation $ (\hat{\boldsymbol{a}},\hat{\boldsymbol{b}},\hat{\boldsymbol{\zeta}}) $ are consistent under Assumption~\ref{c1}-\ref{c2}.
 \end{thm}
 \begin{thm}\label{thm2}
If the estimation $ (\hat{\boldsymbol{a}},\hat{\boldsymbol{b}},\hat{\boldsymbol{\zeta}}) $ are consistent, then they are asymptotically normal with mean centered at the true parameters and variance being the inverse Fisher information matrix.
\end{thm}

The proof of Theorem~\ref{thm1}-\ref{thm2} would be presented in Appendix A.
 \subsection{Bayesian estimation using MCMC method}
Though the MML estimation of $ (\boldsymbol{a},\boldsymbol{b},\boldsymbol{\zeta}) $ are consistent, the other parameters could not be estimated by MML estimation. One natural idea is to estimate the parameters in the proposed model by Monte Carlo Markov Chain (MCMC) method. Actually, MCMC and MML estimation have been already compared in context of IRT \citep{kieftenbeld2012recovery,hendrick2014examination}. It was verified that there were little difference in item parameter recovery between the two methods with samples of 300 or more \citep{kieftenbeld2012recovery}. So MCMC method was eventually employed to estimate the whole parameters in the proposed method.

 We only take the proposed based on probit link as a demonstration, as 2PL IRT model can be very close to 2PNO IRT model through multiplying by a scaling constant 1.702 for the logistic item discrimination parameter \citep{baker2004item}. In details, Gibbs sampling was employed to estimate in unknown parameters in Equation~\eqref{2pno} and Equation~\eqref{pir}. And Metropolis–Hastings algorithm was adopted to estimate $\sigma_{\theta \tau}$ and $\sigma^{2}_{\tau}$.

In order to realize the Gibbs sampling for the 2PNO IRT model, the augmented $Z_{ij}$ was introduced for response variable $Y_{ij}$, where $Z_{ij}\left\vert \theta_{i},a_{j},b_{j},Y_{ij}\right.\sim N\left(  a_{j}\left(  \theta_{i}-b_{j}\right),1\right) $ \citep{albert1992bayesian,albert1993bayesian}. It was assumed that 
\begin{equation}\label{zij}
Y_{ij} =
\begin{dcases}
1&\text{if $Z_{ij}\geq0$},\\
0&\text{if $Z_{ij}<0$}.
\end{dcases}
\end{equation}

 Similarly, the independent random variables $W_{ij}$ were augmented for the missing data indicator $R_{ij}$, which were assumed to follow the normal distribution $ N\left(  \gamma_{0}-\tau_{i}+\zeta_{j}+g\left(  \boldsymbol{R}
_{i,j-1},\gamma_{1}\right)  +l\left(  \boldsymbol{Y}_{ij},\gamma
_{2}\right),1\right) $. So Equation \eqref{pir} could be reformulated as 
\begin{equation}\label{wij}
R_{ij} = 
\begin{dcases}
1&\text{if $W_{ij}\geq0$},\\
0&\text{if $W_{ij}<0$}.
\end{dcases}
\end{equation}

The detailed sampling process would be presented in Appendix B.
\section{Simulation Studies} 
Two simulation studies were conducted to investigate the empirical performance of the proposed model, including parameter recovery and Bayesian model assessment.

\subsection{Simulation \uppercase\expandafter{\romannumeral1}}
 Simulation \uppercase\expandafter{\romannumeral1} is used to compare the parameter recovery of the proposed method with that of listwise deletion. 
\subsubsection{Design}
In the data generation, the number of examinees and the number of items were set to $N=500$ and $J=20$, respectively. The true values of model parameters were set as follows:

$a_{j}\sim\mathrm{Uniform}\left(  0.5,1.5\right),b_{j}\sim\mathrm{N}\left(
	0,1\right),\zeta_{j}\sim\mathrm{N}\left(  0,1\right),j=1,...J. $
	\[
	\left(
	\begin{tabular}
	[c]{l}%
	$\theta_{i}$\\
	$\tau_{i}$%
	\end{tabular}
	\ \right)  \sim N\left(  \left(
	\begin{tabular}
	[c]{l}%
	$0$\\
	$0$%
	\end{tabular}
	\ \right)  ,\left(
	\begin{tabular}
	[c]{ll}%
	$1$ & $\sigma_{\theta \tau}$\\
	$\sigma_{\theta \tau}$ & $1$%
	\end{tabular}
	\ \right)  \right)  ,i=1,...,N.
	\]

As to covariance $\sigma_{\theta \tau}$, also regarded as the correlation between $ \theta $ and $ \tau $ (denoted as $ \rho$ ), were set as 0, 0.4, 0.8 to elucidate the effect of no, a little, large strength of nonignorability. 
		
Equation \eqref{2pno} was used to simulate the dichotomous response data. And then Equation \eqref{pir} was employed to generate the missing responses. Actually, the proportions of missing response is adjusted by $\boldsymbol{\gamma}=(\gamma_{0},\gamma_{1},\gamma_{2}) $. The true values of $\boldsymbol{\gamma}$ and corresponding average missing proportions (denoted by $\overline{p}$) were set as follows:
\begin{equation*}
	\left(
	\begin{tabular}
	[c]{l}%
	$\boldsymbol{\gamma_{0}}$\\
	$\boldsymbol{\gamma_{1}}$\\
	$\boldsymbol{\gamma_{2}}$	\\
	$\boldsymbol{\overline{p}}$
	\end{tabular}
	\right)  = \left(  
	\begin{tabular}{ccccc}
	-2.2&-1.6&-1.1&-0.65&-0.2 \\	
	0.02&0.04&0.04&0.05&0.05 \\	
	-0.2&-0.2&-0.2&-0.25&-0.25\\
	0.096&0.178&0.266&0.371&0.472
	\end{tabular}
	\right)  
\end{equation*}

 100 datasets were simulated for 5 ($\boldsymbol{\gamma}$) $ \times $ 3 ($ \sigma_{\theta\tau} $) =15 conditions. The MCMC sampling procedure was iterated 20,000 and the first 1,5000 iterations were discarded as burn-in. And then the expected a posteriori (EAP) estimation of each parameter can be obtained from its Markov Chain.
\subsubsection{Criteria}
 To assess the performance of parameter recovery, two criteria were applied: mean Bias and mean absolute error (MAE). 

Simply speaking, let $ \boldsymbol{\lambda}=(\lambda_{1},\cdots,\lambda_{K}) $ be a vector of true parameter value. And denote by $ \widehat{\boldsymbol{\lambda_{l}}}=(\widehat{\lambda_{1l}},\cdots,\widehat{\lambda_{Kl}})$ be its EAP estimation, $ l=1,\cdots,L $, where $ L $ is the number of replications. The mean Bias and MAE could be estimated by
\[\overline{Bias}(\boldsymbol{\lambda})=(KL)^{-1}\sum_{l=1}^{L}\sum_{k=1}^{K}(\widehat{\lambda_{kl}}-\lambda_{k}),\]
\[\overline{MAE}(\boldsymbol{\lambda})=(KL)^{-1}\sum_{l=1}^{L}\sum_{k=1}^{K}|\widehat{\lambda_{kl}}-\lambda_{k}|.\]

\subsubsection{Results}
 \begin{table}[htbp]
	
	\scalebox{0.8}[0.8]{
		
		\begin{threeparttable}
			\caption{\label{ignorablebias}The results of parameters recovery for ignorable nonresponses.}
			\begin{tabular}{cccccccc}
				\hline
				\multirow{2}*{$(\gamma_{0},\gamma_{1},\gamma_{2})$}&\multirow{2}*{$ \bar{p} $}&	\multirow{2}*{Parameter}& \multicolumn{2}{c}{ The Proposed Method}  &  & \multicolumn{2}{c}{ Listwise Deletion Method}   \\ \cline{4-5} \cline{7-8}
				&& & Bias& MAE &  & Bias  &MAE \\ \hline
				\multirow{10}*{(-2.2,0.02,-0.2) } &\multirow{10}*{0.093
				}&${\boldsymbol{a}} $	&0.005&0.107&&0.005&0.108\\ 
				&&$ {\boldsymbol{b}} $	&-0.002&0.094&&-0.016&0.095\\ 
				&
				&$ {\boldsymbol{\theta}}$	&0.000&0.292&&0.001&0.292\\ 
				&
				&$ {\boldsymbol{\tau}}$		& -0.003&0.439 &&  \textemdash & \textemdash \\  
				&
				&$ {\boldsymbol{\zeta}}$	& 0.012&0.198&&\textemdash &   \textemdash \\ 
				&
				&	$ \gamma_{0} $	& -0.059&0.166&&\textemdash &   \textemdash \\ 
				&
				&	$ \gamma_{1} $	& 0.024&0.025&&\textemdash &   \textemdash \\ 
				&
				&	$ \gamma_{2} $	& 0.064&0.070&& \textemdash & \textemdash \\  
				&
				&	$\sigma_{\theta\tau}$	 &0.056&0.056&&\textemdash &   \textemdash \\ 
				&
				&	$\sigma^{2}_{\tau} $	
				&-0.087 &0.110&&\textemdash &  \textemdash  \\ \multirow{10}*{(-1.6,0.04,-0.2) } &\multirow{10}*{0.173}&${\boldsymbol{a}} $	&0.015&0.109&&0.016&0.111\\ 
				&&$ {\boldsymbol{b}} $	&-0.010&0.101&&-0.028&0.105\\ 
				&
				&$ {\boldsymbol{\theta}}$	&0.004&0.309&&0.005&0.310\\ 
				&
				&$ {\boldsymbol{\tau}}$		& 0.001&0.370&&\textemdash &  \textemdash  \\  
				&
				&$ {\boldsymbol{\zeta}}$	& 0.036&0.208& &  \textemdash & \textemdash \\ 
				&
				&	$ \gamma_{0} $	& -0.095&0.192&&\textemdash &  \textemdash  \\ 
				&
				&	$ \gamma_{1} $	& 0.010&0.017&&\textemdash &  \textemdash  \\ 
				&
				&	$ \gamma_{2} $	& 0.103&0.103&&  \textemdash & \textemdash \\  
				&
				&	$\sigma_{\theta\tau}$	& 0.059&0.059&&  \textemdash & \textemdash \\ 
				&
				&	$\sigma^{2}_{\tau} $	
				&-0.033 &0.081&&  \textemdash & \textemdash \\ \multirow{10}*{(-1.1,0.04,-0.2) } &\multirow{10}*{0.264}& ${\boldsymbol{a}} $	&0.009&0.116&&0.009&0.116\\ 
				&&$ {\boldsymbol{b}} $	&-0.017&0.112&&-0.041&0.117\\ 
				&
				&$ {\boldsymbol{\theta}}$	&0.010&0.331&&0.011&0.333\\ 
				&
				&$ {\boldsymbol{\tau}}$		& 0.002&0.326&&\textemdash &  \textemdash \\  
				&
				&$ {\boldsymbol{\zeta}}$	& 0.008&0.195&&\textemdash &  \textemdash \\ 
				&
				&	$ \gamma_{0} $	& -0.079&0.190&&\textemdash &  \textemdash \\ 
				&
				&	$ \gamma_{1} $	& 0.008&0.015&&\textemdash &  \textemdash  \\ 
				&
				&	$ \gamma_{2} $	& 0.127&0.127&&\textemdash &  \textemdash  \\  
				&
				&	$\sigma_{\theta\tau}$	& 0.053&0.053&&\textemdash &  \textemdash  \\ 
				&
				&	$\sigma^{2}_{\tau} $	
				&-0.040 &0.089&&\textemdash &  \textemdash \\ \multirow{10}*{(-0.65,0.05,-0.25) } &\multirow{10}*{0.370}&${\boldsymbol{a}} $	&0.017&0.130&&0.017&0.130\\ 
				&&$ {\boldsymbol{b}} $	&-0.063&0.133&&-0.091&0.145\\ 
				&
				&$ {\boldsymbol{\theta}}$	&0.003&0.361&&0.005&0.362\\ 
				&
				&$ {\boldsymbol{\tau}}$		& 0.004&0.297&&\textemdash &  \textemdash \\  
				&
				&$ {\boldsymbol{\zeta}}$	& -0.001&0.191&&\textemdash &  \textemdash \\ 
				&
				&	$ \gamma_{0} $	& -0.104&0.193&& \textemdash & \textemdash \\ 
				&
				&	$ \gamma_{1} $	& 0.006&0.015&&  \textemdash & \textemdash \\ 
				&
				&	$ \gamma_{2} $	& 0.188&0.188&&  \textemdash & \textemdash \\  
				&
				&	$\sigma_{\theta\tau}$	& 0.050&0.050& &  \textemdash & \textemdash \\ 
				&
				&	$\sigma^{2}_{\tau} $	
				&-0.046 &0.091& &  \textemdash & \textemdash \\\multirow{10}*{(-0.2,0.05,-0.25) } &\multirow{10}*{0.463}& ${\boldsymbol{a}} $	&0.018&0.146&&0.020&0.146\\ 
				&&$ {\boldsymbol{b}} $	&-0.074&0.148&&-0.109&0.164\\ 
				&
				&$ {\boldsymbol{\theta}}$	&0.007&0.398&&0.009&0.399\\ 
				&
				&$ {\boldsymbol{\tau}}$		& -0.003&0.287&&\textemdash &  \textemdash \\  
				&
				&$ {\boldsymbol{\zeta}}$	& 0.047&0.201&&\textemdash &  \textemdash  \\ 
				&
				&	$ \gamma_{0} $	& -0.180&0.220&&\textemdash &  \textemdash \\ 
				&
				&	$ \gamma_{1} $	& 0.006&0.012&&  \textemdash & \textemdash \\ 
				&
				&	$ \gamma_{2} $	& 0.198&0.198&&\textemdash &  \textemdash \\  
				&
				&	$\sigma_{\theta\tau}$	& 0.056&0.056&&\textemdash &  \textemdash \\ 
				&
				&	$\sigma^{2}_{\tau} $	
				&-0.047 &0.084&&\textemdash &  \textemdash \\ 
				\hline
			\end{tabular} 
			\small \textit{Note}. Bias and MAE refer to the mean Bias and MAE for each parameter in the third column, respectively.	
	\end{threeparttable}}
\end{table}

\begin{table}[htbp]
	
	\scalebox{0.8}[0.8]{
		
		\begin{threeparttable}
			\caption{\label{nonignorablebias0.4}The results of parameters recovery for nonignorable nonresponses ($\rho=0.4$).}
			\begin{tabular}{cccccccc}
				\hline
				\multirow{2}*{$(\gamma_{0},\gamma_{1},\gamma_{2})$}&\multirow{2}*{$ \bar{p} $}&	\multirow{2}*{Parameter}& \multicolumn{2}{c}{ The Proposed Method}  &  & \multicolumn{2}{c}{ Listwise Deletion Method}   \\ \cline{4-5} \cline{7-8}
				&& & Bias& MAE &  & Bias  &MAE \\ \hline
				\multirow{10}*{(-2.2,0.02,-0.2) } &\multirow{10}*{0.098
				}&${\boldsymbol{a}} $	&0.013&0.106&&0.004&0.106\\ 
				&&$ {\boldsymbol{b}} $	&0.003&0.098&&-0.023&0.103\\ 
				&
				&$ {\boldsymbol{\theta}}$	&0.008&0.289&&0.007&0.295\\ 
				&
				&$ {\boldsymbol{\tau}}$		&0.003 &0.421&&\textemdash &  \textemdash  \\  
				&
				&$ {\boldsymbol{\zeta}}$	&0.009 &0.228&&\textemdash &  \textemdash  \\ 
				&
				&	$ \gamma_{0} $	& -0.041&0.189&& \textemdash & \textemdash \\ 
				&
				&	$ \gamma_{1} $	& 0.020&0.021& &  \textemdash & \textemdash \\ 
				&
				&	$ \gamma_{2} $	& 0.057&0.070& &  \textemdash & \textemdash \\  
				&
				&	$\sigma_{\theta\tau}$	&-0.014 &0.048& &  \textemdash & \textemdash \\ 
				&
				&	$\sigma^{2}_{\tau} $	
				&-0.051 &0.096& &  \textemdash & \textemdash \\ \multirow{10}*{(-1.6,0.04,-0.2) } &\multirow{10}*{0.178}&${\boldsymbol{a}} $	&0.006&0.112&&-0.005&0.114\\ 
				&&$ {\boldsymbol{b}} $	&-0.016&0.106&&-0.055&0.118\\ 
				&
				&$ {\boldsymbol{\theta}}$	&0.005&0.306&&0.005&0.316\\ 
				&
				&$ {\boldsymbol{\tau}}$		& 0.008&0.357& &  \textemdash & \textemdash \\  
				&
				&$ {\boldsymbol{\zeta}}$	&0.024 &0.204&&\textemdash  & \textemdash \\ 
				&
				&	$ \gamma_{0} $	& -0.067&0.199&&  \textemdash & \textemdash \\ 
				&
				&	$ \gamma_{1} $	& 0.004&0.015&&\textemdash &  \textemdash\\ 
				&
				&	$ \gamma_{2} $	& 0.103&0.103& &  \textemdash & \textemdash \\  
				&
				&	$\sigma_{\theta\tau}$	&0.010 &0.044& &  \textemdash & \textemdash \\ 
				&
				&	$\sigma^{2}_{\tau} $	
				& 0.012&0.100&&\textemdash &  \textemdash  \\ \multirow{10}*{(-1.1,0.04,-0.2) } &\multirow{10}*{0.266}& ${\boldsymbol{a}} $	&0.010&0.127&&-0.010&0.124\\ 
				&&$ {\boldsymbol{b}} $	&-0.034&0.117&&-0.091&0.141\\ 
				&
				&$ {\boldsymbol{\theta}}$	&0.004&0.325&&0.006&0.339\\ 
				&
				&$ {\boldsymbol{\tau}}$		&-0.007 &0.316& &  \textemdash & \textemdash \\  
				&
				&$ {\boldsymbol{\zeta}}$	&-0.005 &0.165&&\textemdash &  \textemdash  \\ 
				&
				&	$ \gamma_{0} $	& -0.075&0.159&& \textemdash & \textemdash \\ 
				&
				&	$ \gamma_{1} $	& 0.005&0.014&&\textemdash &  \textemdash \\ 
				&
				&	$ \gamma_{2} $	& 0.130&0.130& &  \textemdash & \textemdash \\  
				&
				&	$\sigma_{\theta\tau}$	& 0.014&0.051& &  \textemdash & \textemdash \\ 
				&
				&	$\sigma^{2}_{\tau} $	
				& 0.004&0.088&&\textemdash & \textemdash \\ \multirow{10}*{(-0.65,0.05,-0.25) } &\multirow{10}*{0.371}&${\boldsymbol{a}} $	&0.008&0.133&&-0.008&0.135\\ 
				&&$ {\boldsymbol{b}} $	&-0.069&0.139&&-0.151&0.189\\ 
				&
				&$ {\boldsymbol{\theta}}$	&0.006&0.355&&0.009&0.377\\ 
				&
				&$ {\boldsymbol{\tau}}$		& 0.002&0.296&&\textemdash &  \textemdash  \\  
				&
				& $ {\boldsymbol{\zeta}}$	& -0.020&0.187&&  \textemdash & \textemdash \\ 
				&
				&	$ \gamma_{0} $	& -0.075&0.193& &  \textemdash & \textemdash \\ 
				&
				&	$ \gamma_{1} $	& 0.002&0.014& &  \textemdash & \textemdash \\ 
				&
				&	$ \gamma_{2} $	& 0.190&0.190& &  \textemdash & \textemdash \\  
				&
				&	$\sigma_{\theta\tau}$	& 0.009&0.053& &  \textemdash & \textemdash \\ 
				&
				&	$\sigma^{2}_{\tau} $	
				&0.041 &0.103& &  \textemdash & \textemdash \\\multirow{10}*{(-0.2,0.05,-0.25) } &\multirow{10}*{0.480}& ${\boldsymbol{a}} $	&-0.003&0.153&&-0.018&0.153\\ 
				&&$ {\boldsymbol{b}} $	&-0.123&0.177&&-0.231&0.255\\ 
				&
				&$ {\boldsymbol{\theta}}$	&0.003&0.386&&0.009&0.419\\ 
				&
				&$ {\boldsymbol{\tau}}$		& -0.002&0.286& &  \textemdash & \textemdash \\  
				&
				&$ {\boldsymbol{\zeta}}$	& -0.021&0.189&&\textemdash &  \textemdash  \\ 
				&
				&	$ \gamma_{0} $	& -0.110&0.171&&  \textemdash & \textemdash \\ 
				&
				&	$ \gamma_{1} $	& 0.006&0.014&&  \textemdash & \textemdash \\ 
				&
				&	$ \gamma_{2} $	& 0.203&0.203& &  \textemdash & \textemdash \\  
				&
				&	$\sigma_{\theta\tau}$	& -0.011&0.048& &  \textemdash & \textemdash \\ 
				&
				&	$\sigma^{2}_{\tau} $	
				&0.004 &0.093& &  \textemdash & \textemdash \\ 
				\hline
			\end{tabular} 
			\small \textit{Note}. Bias and MAE refer to the mean Bias and MAE for each parameter in the third column, respectively.
	\end{threeparttable}}
\end{table}

\begin{table}[htbp]
	
	\scalebox{0.8}[0.8]{
		
		\begin{threeparttable}
			\caption{\label{nonignorablebias0.8}The results of parameters recovery for nonignorable nonresponses ($ \rho=0.8 $).}
			\begin{tabular}{cccccccc}
				\hline
				\multirow{2}*{$(\gamma_{0},\gamma_{1},\gamma_{2})$}&\multirow{2}*{$ \bar{p} $}&	\multirow{2}*{Parameter}& \multicolumn{2}{c}{ The Proposed Method}  &  & \multicolumn{2}{c}{ Listwise Deletion Method}   \\ \cline{4-5} \cline{7-8}
				&& & Bias& MAE &  & Bias  &MAE \\ \hline
				\multirow{10}*{(-2.2,0.02,-0.2) } &\multirow{10}*{0.097
				}&${\boldsymbol{a}} $	&0.011&0.105&&-0.015&0.110\\ 
				&&$ {\boldsymbol{b}} $	&-0.009&0.104&&-0.051&0.120\\ 
				&
				&$ {\boldsymbol{\theta}}$	&0.001&0.278&&-0.002&0.303\\ 
				&
				&$ {\boldsymbol{\tau}}$		& 0.001&0.373& &  \textemdash & \textemdash \\  
				&
				&$ {\boldsymbol{\zeta}}$	&0.026 &0.206& &  \textemdash & \textemdash \\ 
				&
				&	$ \gamma_{0} $	& -0.053&0.163& &  \textemdash & \textemdash \\ 
				&
				&	$ \gamma_{1} $	& 0.024&0.024&&  \textemdash & \textemdash \\ 
				&
				&	$ \gamma_{2} $	& 0.038&0.079& &  \textemdash & \textemdash \\  
				&
				&	$\sigma_{\theta\tau}$	& -0.020&0.062& &  \textemdash & \textemdash \\ 
				&
				&	$\sigma^{2}_{\tau} $	
				& -0.050&0.122& &  \textemdash & \textemdash \\ \multirow{10}*{(-1.6,0.04,-0.2) } &\multirow{10}*{0.180}&${\boldsymbol{a}} $	&0.006&0.116&&-0.032&0.126\\ 
				&&$ {\boldsymbol{b}} $	&-0.018&0.110&&-0.086&0.140\\ 
				&
				&$ {\boldsymbol{\theta}}$	&0.004&0.283&&0.005&0.323\\ 
				&
				&$ {\boldsymbol{\tau}}$		& 0.002&0.328&&\textemdash &  \textemdash \\  
				&
				&$ {\boldsymbol{\zeta}}$	& 0.008&0.205& &  \textemdash & \textemdash \\ 
				&
				&	$ \gamma_{0} $	& -0.058&0.196& &  \textemdash & \textemdash \\ 
				&
				&	$ \gamma_{1} $	& 0.008&0.017&&  \textemdash & \textemdash \\ 
				&
				&	$ \gamma_{2} $	& 0.095&0.111& &  \textemdash & \textemdash \\  
				&
				&	$\sigma_{\theta\tau}$	&0.000 &0.062& &  \textemdash & \textemdash \\ 
				&
				&	$\sigma^{2}_{\tau} $	
				&0.021 &0.130& &  \textemdash & \textemdash \\ \multirow{10}*{(-1.1,0.04,-0.2) } &\multirow{10}*{0.267}& ${\boldsymbol{a}} $	&-0.002&0.122&&-0.057&0.137\\ 
				&&$ {\boldsymbol{b}} $	&-0.033&0.124&&-0.139&0.185\\ 
				&
				&$ {\boldsymbol{\theta}}$	&0.003&0.298&&0.006&0.361\\ 
				&
				&$ {\boldsymbol{\tau}}$		&0.003 &0.304& &  \textemdash & \textemdash \\  
				&
				&$ {\boldsymbol{\zeta}}$	& 0.017&0.200& &  \textemdash & \textemdash \\ 
				&
				&	$ \gamma_{0} $	& -0.094&0.196&&  \textemdash & \textemdash \\ 
				&
				&	$ \gamma_{1} $	& 0.008&0.017& &  \textemdash & \textemdash \\ 
				&
				&	$ \gamma_{2} $	& 0.119&0.136& &  \textemdash & \textemdash \\  
				&
				&	$\sigma_{\theta\tau}$	& -0.003&0.063& &  \textemdash & \textemdash \\ 
				&
				&	$\sigma^{2}_{\tau} $	
				&0.019 &0.120& &  \textemdash & \textemdash \\ \multirow{10}*{(-0.65,0.05,-0.25) } &\multirow{10}*{0.365}&${\boldsymbol{a}} $	&-0.013&0.142&&-0.078&0.161\\ 
				&&$ {\boldsymbol{b}} $	&-0.085&0.161&&-0.247&0.276\\ 
				&
				&$ {\boldsymbol{\theta}}$	&0.007&0.314&&0.011&0.407\\ 
				&
				&$ {\boldsymbol{\tau}}$		& 0.005&0.284& &  \textemdash & \textemdash \\  
				&
				&$ {\boldsymbol{\zeta}}$	& 0.022&0.195& &  \textemdash & \textemdash \\ 
				&
				&	$ \gamma_{0} $	& -0.117&0.204& &  \textemdash & \textemdash \\ 
				&
				&	$ \gamma_{1} $	& 0.002&0.014&&  \textemdash & \textemdash \\ 
				&
				&	$ \gamma_{2} $	& 0.192&0.193& &  \textemdash & \textemdash \\  
				&
				&	$\sigma_{\theta\tau}$	&0.025 &0.053& &  \textemdash & \textemdash \\ 
				&
				&	$\sigma^{2}_{\tau} $	
				& 0.077&0.128& &  \textemdash & \textemdash \\\multirow{10}*{(-0.2,0.05,-0.25) } &\multirow{10}*{0.464}& ${\boldsymbol{a}} $	&-0.018&0.158&&-0.082&0.181\\ 
				&&$ {\boldsymbol{b}} $	&-0.110&0.183&&-0.324&0.347\\ 
				&
				&$ {\boldsymbol{\theta}}$	&0.185&0.335&&0.011&0.455\\ 
				&
				&$ {\boldsymbol{\tau}}$		& 0.004&0.281&&\textemdash &  \textemdash  \\  
				&
				&$ {\boldsymbol{\zeta}}$	&0.052 &0.207& &  \textemdash & \textemdash \\ 
				&
				&	$ \gamma_{0} $	& -0.180&0.216& &  \textemdash & \textemdash \\ 
				&
				&	$ \gamma_{1} $	& 0.009&0.017&& \textemdash & \textemdash \\ 
				&
				&	$ \gamma_{2} $	& 0.202&0.202&&\textemdash &  \textemdash \\  
				&
				&	$\sigma_{\theta\tau}$	& 0.003&0.064& &  \textemdash & \textemdash \\ 
				&
				&	$\sigma^{2}_{\tau} $	
				&0.043 &0.132& &  \textemdash & \textemdash \\ 
				\hline
			\end{tabular} 
			\small \textit{Note}. Bias and MAE refer to the mean Bias and MAE for each parameter in the third column, respectively.
	\end{threeparttable}}
\end{table}
Table~\ref{ignorablebias} presents the results of parameter recovery under both the proposed method and listwise deletion for the ignorable missingness ($\rho=0$). The results show that the parameter recovery under the two methods are nearly similar for $ \boldsymbol{a} $, $ \boldsymbol{b} $ and $ \boldsymbol{\theta} $. That is, under the two methods, the item parameters $ \boldsymbol{a} $ and $ \boldsymbol{b} $ are well recovered, as their biases are close to 0 and MAEs are around 0.1. Relatively higher MAE for the person ability parameter $ \boldsymbol{\theta} $ is present. And the bias of $ \boldsymbol{\theta} $ is very close to 0. For the other parameters in the proposed method, they are almost well recovered, except the person missingness parameter $ \boldsymbol{\tau} $. One explanation might simply be that the dimensions of person parameters $ \boldsymbol{\theta} $ and $ \boldsymbol{\tau} $ are much higher than the item parameters.

Table~\ref{nonignorablebias0.4} and Table~\ref{nonignorablebias0.8} show the results of parameters recovery under little ($\rho=0.4$) and large ($\rho=0.8$) nonignorable missingness, respectively. Generally speaking, the recovery of parameters under the proposed method is better than applying listwise deletion. At the same time, as the missing proportion is higher and nonignorability is stronger, the superiority is more obvious. In details, for the proposed method, the item parameters $ \boldsymbol{a} $ and $ \boldsymbol{b} $ are always well recovered across all 5 ($\boldsymbol{\gamma}$) $ \times $ 2 ($ \sigma_{\theta\tau} $) =10 conditions as the bias is very close to 0 and the MAE is no more than 0.2. Note that for the fixed $\rho$, the MAE of $\boldsymbol{\tau}$ decreases with the increasing of missing proportions. That is, $\boldsymbol{\tau}$ recovers better when the missing responses is more, which also confirms that the missingness could be attributed to latent person missing trait.

\subsection{Simulation \uppercase\expandafter{\romannumeral2}}
Besides evaluating the parameter recovery, we are also interested in assessing whether the missingness is ignorable or nonignorable. Simulation \uppercase\expandafter{\romannumeral2} was conducted to assess whether the missingness is ignorable or nonignorable.

\subsubsection{Design}
 The settings of true model parameters are the same as in Simulation \uppercase\expandafter{\romannumeral1}, except parameter $\rho$. More specifically, $ \rho $ is only to confirm whether the missingness is nonignorable or ignorable in this simulation. That is, in this simulation, the nonignorable ($ \rho\neq0 $) and ignorable ($ \rho=0 $) models were employed to fit the data generate from the nonignorable model ($ \rho=0.8 $). 
\subsubsection{Criteria}
Within Bayesian framework, there are some commonly used 
criteria for model selection. In this section, these criteria were only applied to the distribution of $ \boldsymbol{R}|\boldsymbol{Y} $, because we only focus on the missing data mechanism. 

One criterion to evaluate the model fit is deviance information criterion (DIC; \citealp{spiegelhalter2002bayesian}). This criterion takes into account the trade-off relationship between the adequacy of model fitting and the number of model parameters. The model with a smaller DIC value fits the data better.

The other criterion to compare the two models in terms of fitting is the logarithm of the
pseudo marginal likelihood (LPML; \citealp{geisser1979predictive, ibrahim2001bayesian}). 
The model with a larger LPML has a better fit of the data.

The detailed computational processes of the two criteria would be presented in Appendix C.

\subsubsection{Results}
\begin{figure}[htb]
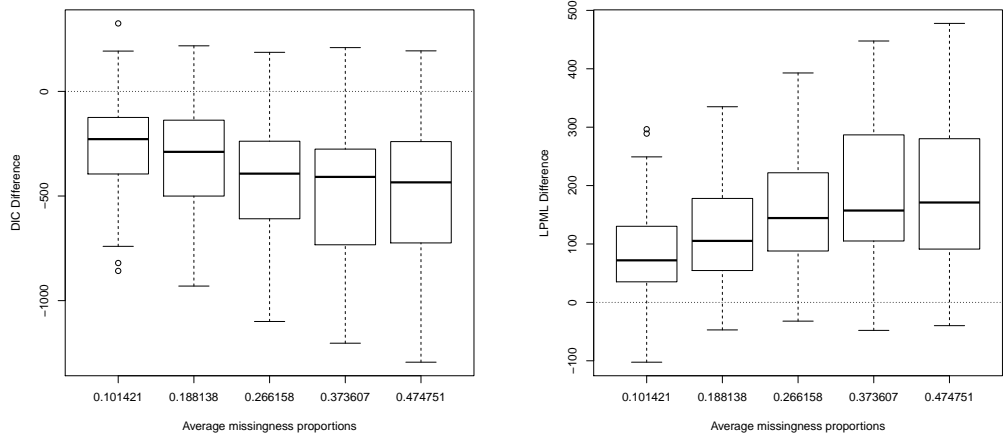

	\begin{center}
		\begin{tabular}{cc}
			\includegraphics[width=2.6in, ]{DIC.png} & \includegraphics[width=2.6in]{LPML.png}
		\end{tabular}
	\end{center}
	\caption{Boxplots of DIC and LPML difference between nonignorable and ignorable model.\label{DICANDLPML}}	
\end{figure}

The DIC and LPML difference between the nonignorable and ignorable model were calculated and presented by boxplots in Figure~\ref{DICANDLPML}. 

The boxes of DIC difference between nonignorable and ignorable model are always below 0, indicating that the nonignorable model fits the data better. Meanwhile, the LPML differences between nonignorable and ignorable model are always more than 0, which shows the same conclusion as DIC. Furthermore, the DIC and LPML difference increase with the proportions of missing responses. So the nonignorable model has more distinct advantages in terms of model selection when the missing proportion is higher. 

\section{Analysis of the PISA data}
The PISA is an international education assessment that measures students' skills and knowledge in science, mathematics, reading, and so on. Its data set is available and free on  {\url{http://www.oecd.org/pisa/data/}}.
 
 In this section, PISA data set was employed to illustrate the detailed use of the proposed method and further interpret the parameters.
\subsection{Data set}
In this study, the data set is chosen from a science subtest in the 2015 computer-based PISA in Dominican Republic. The invalid and not applicable samples are excluded from the data set. The valid sample size is 493, in which 173 individuals reached all 17 items. The overall missing proportion of the dataset is $ 22.9\% $ ($ 9.1\% $ omitted items and $ 13.8\% $ not-reached items). Only DS465Q01C is scored polytomously with 0 (no credit), 1 (partial credit), and 2 (full credit) scores. As the proposed method is focused on dichotomous responses, only full credit is treated as a correct response, while the other two score categories are treated as incorrect responses.

\subsection{Analysis}
The nonignorable and ignorable models under the proposed method were used to fit the data. DIC and LPML of $ \boldsymbol{R}|\boldsymbol{Y} $ under the both models were computed for the purpose of model selection. The DICs under nonignorable and ignorable model are 5504 and 5857, respectively. And the LPMLs under nonignorable and ignorable model are -2895 and -3016, respectively.
Judging by the both criteria, nonignorable model was selected. So the following analysis based on the proposed method refers to the nonignorable models.

\begin{figure}[htb] 	
		\centering	
		\includegraphics[width=4.5in]{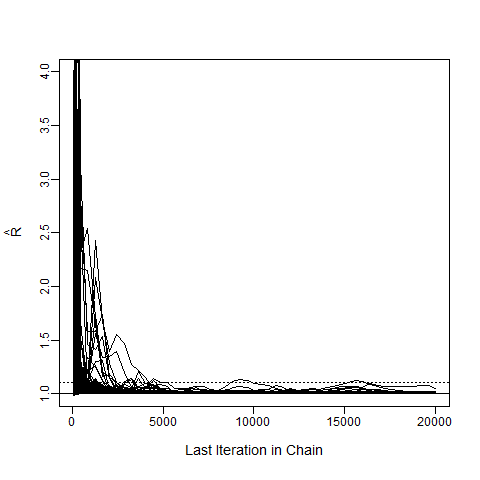}
		\caption{\label{rhat}The trace plots of $\hat{R}$. \textit{Note}. The dashed line is 1.1.} 
\end{figure}

Three Markov chains started at over dispersed starting values were used and each chain had 20000 iterations. The Gelman-Rubin convergence statistic 
$\hat{R}$
 \citep{brooks1998general} was computed to assess the convergences of all parameters in the proposed model. Convergence is evaluated by the value of $\hat{R}$. That is, if the $\hat{R}$ is less than 1.1, the parameter achieves convergence. The values of $\hat{R}$ can be obtained based on the
\textquotedblleft coda\textquotedblright\ R package \citep{plummer2006coda}. The trace plots of $\hat{R}$ for all parameters in the proposed method are presented in Figure~\ref{rhat}. It suggests that $\hat{R}$ is generally less than 1.1 after
5,000 iterations for each parameter, indicating the perfect convergence of all the model parameters. 

Similar to Simulation \uppercase\expandafter{\romannumeral1}, listwise deletion was also employed to get the comparative analysis of parameter estimate. That is, 173 samples with complete responses were used to get the parameter estimates based on traditional IRT models. 

\subsection{Results}
Table~\ref{real} summarized the missing proportions of each item and the results of item parameters estimation based on the proposed method. For DS465Q01C and DS438Q03C, their EAP of $\zeta$ is much higher than the other items, which means the two items are more likely to be omitted. For the last three items, the proportions of missing are higher, despite their $\zeta$ are estimated to be negative. One possible explanation is that the missing data mechanism for the three items is mainly the effect that previous missing items bring on current item, so the missingness for these items may be largely not-reached.
In general, the items with higher $\zeta$ or at the end of test tend to miss.
\begin{table}[htbp]
	\begin{threeparttable}
		\caption{Item parameters estimation for the PISA subtest data based on the proposed method.}
		\label{real}
		\begin{tabular}{cccccccccc}
			\hline
			\multirow{2}*{Item}& \multirow{2}*{$ p $}&\multicolumn{2}{c}{ $ a $}   &&\multicolumn{2}{c}{ $ b $}&&\multicolumn{2}{c}{$  \zeta $}\\
			\cline{3-4} \cline{6-7} \cline{9-10}
			
			& &$ EAP $ & $ SE$ & &$ EAP $ & $ SE$ & &$EAP$ &$SE$  \\ \hline
			DS465Q01C&0.296 &1.078 & 0.003 && 2.302 & 0.004 && 0.784 & 0.001 \\ 
			CS465Q02S&0.142 & 0.329& 0.001& & 0.743 &0.004 &  &-0.023  &  0.001\\ 
			CS465Q04S& 0.123&0.177 & 0.000&  &3.150  &  0.008&&  -0.225 & 0.001 \\ 
			DS131Q02C&0.203&0.730&0.002&& 2.615&0.005& &0.215&0.001\\
			DS131Q04C&0.237&1.099&0.003& &1.720&0.003&& 0.333&0.001\\
			CS428Q01S&0.097&0.867&0.002& &1.182&0.002& &-0.748&0.002\\
			CS428Q03S&0.102&0.827&0.002&& 1.102&0.002& &-0.741&0.002\\
			DS428Q05C&0.275&1.218&0.004& &2.243&0.004& &0.418&0.001\\
			DS514Q02C&0.247&0.781&0.002&& 0.019&0.002&& 0.191&0.001\\
			DS514Q03C&0.169&0.784&0.002& &0.584&0.002& &-0.355&0.001\\
			DS514Q04C&0.298&1.535&0.003& &1.689&0.002&& 0.342&0.001\\
			CS438Q01S&0.197&0.711&0.002&& 0.584&0.002& &-0.355&0.001\\
			CS438Q02S&0.239&0.558&0.001& &0.872&0.003& &-0.125&0.001\\
			DS438Q03C&0.459&1.045&0.003& &2.347&0.004&& 0.965&0.001\\
			CS415Q07S&0.247&0.503&0.001&& 0.087&0.002&& -0.277&0.001\\
			CS415Q02S&0.279&0.709&0.002& &0.243&0.002& &-0.123&0.001\\
			CS415Q08S&0.275&0.546&0.002&& 1.344&0.004&& -0.228&0.005\\
			\hline
		\end{tabular}
		\small \textit{Note}. $p$ refers to the missing proportion for each item. $ EAP $=expected a posteriori, $ SE $=standard error.
	\end{threeparttable}
	\begin{center}
		\begin{threeparttable}
			\caption{Item parameters estimation for the PISA subtest data based on listwise deletion.}
			\label{realirt}
			\begin{tabular}{ccccccc}
				\hline
				\multirow{2}*{Item}&  &\multicolumn{2}{c}{ $ a $}   &&\multicolumn{2}{c}{ $ b $}\\
				\cline{3-4} \cline{6-7}
				&  &$ EAP $ & $ SE$ & &$ EAP $ & $ SE$  \\ \hline
				DS465Q01C&&1.624&0.006&&1.757& 0.003\\ 
				CS465Q02S&&0.500&0.002&&0.368& 0.004\\ 
				CS465Q04S&&0.259&0.001&&2.165&  0.008\\ 
				DS131Q02C&&0.904&0.003&&1.964&0.005 \\
				DS131Q04C&&1.477&0.005&&1.186&0.002 \\
				CS428Q01S&&1.118&0.003&&0.840& 0.002\\
				CS428Q03S&&0.834&0.003&&0.739&0.003 \\
				DS428Q05C&&1.298&0.005&&1.755&0.004 \\
				DS514Q02C&&0.627&0.002&&-0.570&0.003 \\
				DS514Q03C&&0.801&0.003&&1.763& 0.005\\
				DS514Q04C&&1.629&0.005&&1.203& 0.002\\
				CS438Q01S&&0.729&0.002&&-0.246&0.002 \\
				CS438Q02S&&0.771&0.002&&0.469& 0.002\\
				DS438Q03C&&1.224&0.004&&1.950& 0.005\\
				CS415Q07S&&0.427&0.002&&-0.526& 0.004\\
				CS415Q02S&&0.694&0.002&&-0.116& 0.002\\
				CS415Q08S&&0.465&0.002&&0.962&0.005 \\
				\hline
			\end{tabular}
			\small \textit{Note}. $ EAP $=expected a posteriori, $ SE $=standard error.
		\end{threeparttable} 
	\end{center} 
\end{table}

The results of item parameter based on listwise deletion are presented in Table~\ref{realirt}. Obviously, for the same parameter, there is a big difference for the two methods. Specifically, the EAPs of item difficulty parameters in Table~\ref{real} are always higher that in Table~\ref{realirt}. A likely cause is that the missing responses often occur for the examinees with low ability and their responses are deleted in the listwise deletion leading to the underestimates of item difficulty parameters. 

\begin{figure}[htp]
	\centering
		\begin{tabular}{cc}
			\includegraphics[width=2.75in, ]{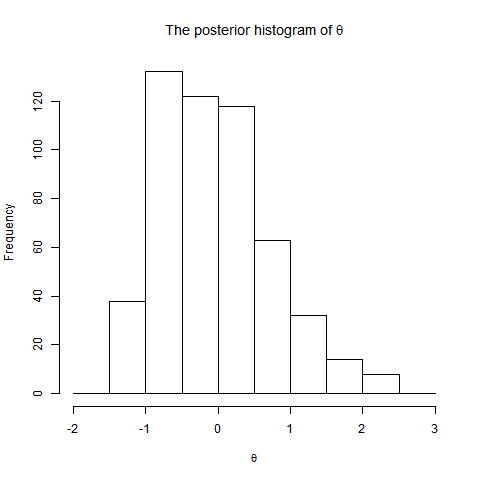} & \includegraphics[width=2.75in]{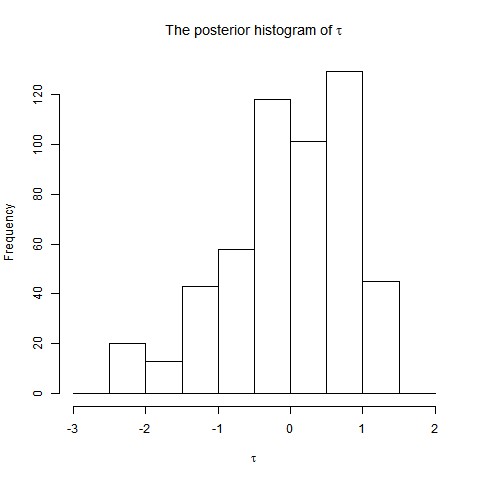}
		\end{tabular}
	\caption{\label{hist}The posterior histograms of person parameters based on the proposed method.} 
	\includegraphics[width=2.75in]{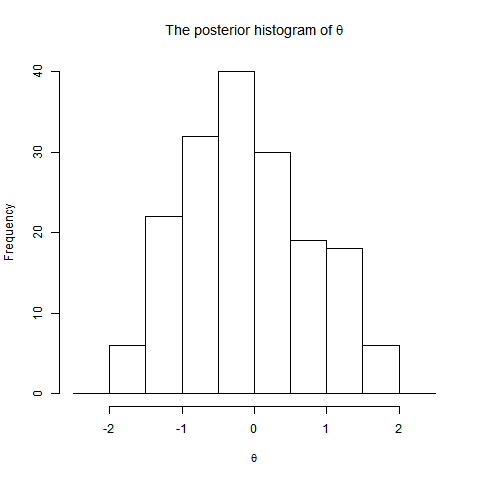}
\caption{\label{histtheta} The posterior histograms of person parameters based on listwise deletion.} 
\end{figure}

The posterior histograms of person parameters in the proposed method are shown in Figure~\ref{hist}, and the posterior histogram of ability parameters based on listwise deletion is presented in Figure~\ref{histtheta}. For the posterior histograms of ability parameters, there also exits a big difference between the two figures. The posterior histograms of ability parameters in Figure~\ref{hist} is sharper than that in Figure~\ref{histtheta}, indicating the variance of the former is smaller.

The results of parameter estimation for other parameters in the proposed model are present in Table~\ref{other}. $\sigma_{\theta\tau}$ and $\sigma^{2}_{\tau} $ are estimated at 0.405 and 1.000, respectively. So the correlation between $\theta$ and $\tau$ is about 0.405, which is significantly more than 0. This also confirms that the missingness is nonignorable, which is consistent with the result of Bayesian model selection. And the EAP of $ \gamma_{1} $ is 0.204 and more than 0 obviously, showing that the term $g\left(  \boldsymbol{R}_{i,j-1}
,\boldsymbol{\gamma}_{1}\right)  =\gamma_{1}\sum\limits_{h=2}^{j-1}%
{R}_{i,h-1}$ in Equation \eqref{pir} is reasonable and necessary.  
	\begin{table}[htb]\caption{The EAP and SE of the other parameters based on the proposed method.}
	\label{other}
	\begin{center}
		\begin{threeparttable} 
			\begin{tabular}{cccccccccc}
				\hline
				Parameter	& $\sigma_{\theta\tau}$ && $\sigma^{2}_{\tau} $& & $ \gamma_{0} $ && $ \gamma_{1} $ &&$ \gamma_{2} $  \\ \hline
				$ EAP $	& 0.405 && 1.000& & -1.53& & 0.204 && -0.034 \\ 
				$ SE $	& 0.001 && 0.002& & 0.001 && 0.000 &&  0.000\\ \hline
			\end{tabular} 
			\small \textit{Note}. $ EAP $=expected a posteriori, $ SE $=standard error.
		\end{threeparttable} 
	\end{center}
\end{table} 
\section{Discussion}
This paper proposed an IRT-based method to model omitted and not-reached items based on SLM. It was proved that the item parameter estimate based on MML estimation is consistent. Further more, MCMC methods was given to estimate all the model parameters based on Probit link. Bayesian model selection for nonignorable and ignorable model was explored using DIC and LPML criteria. The usage and the performance of the proposed method were demonstrated by using the 2015 PISA computer-based science subtest data as an example. The real data analysis indicated that the missingness is nonignorable, as the correlation between $\theta$ and $\tau$ is significantly greater than 0. And both the DIC and LPML also gave the same conclusion. 

The proposed method could handle both ignorable and nonignorable missingness by adjusting the  correlation between person missing parameter and ability. If the correlation is more than 0, $ \pi_{ij} $ in Equation \eqref{pir} depends on the ability $\theta_{i}$, so that the missingness is nonignorable (NMAR). If the correlation is 0, $ \pi_{ij} $ only depends on the observed response $\tau_{i}$  and $\zeta_{j}$, as well as cumulative number of missing response. Therefore, the missingness is ignorable. 

Despite such promising results, other issues should be investigated in the future. The first one is whether the proposed method could be applied to estimate more complicated IRT models, such as the partial credit model \citep{masters1982rasch}, the generalized partial credit model \citep{muraki1992generalized}. Second, if the last item at the end of test is missed and the penultimate one is observed, the last nonresponse may occur due to skipped item or the time limit. It is still unknown how to clarify whether the last nonresponse is omitted or not-reached. In this case, maybe response time could be taken into account to get further information. Finally, the missingness not only occurs in item response data but also response time data for computer based test. It is still a promising issue to extend the proposed method to the mixture model for response times and response accuracy \citep{wang2015mixture}.

\section*{Acknowledgments}
This work is supported by the National Natural Science Foundation of China (grant number 11571069).

\section*{Appendix A: Proof of theorems}
\paragraph*{Proof of Theorem~\ref{thm1}}
Since the items
are assumed to be independent in context of IRT, we only prove the consistency results for the fixed item. For notational convenience, the item subscript $j$ was omitted in the following proof. Moreover, define $\boldsymbol{\lambda}=(a,b,\zeta)'$ as the item parameters vector for the item and $\boldsymbol{\lambda}_{0}$ as its true value.
Actually, the parameters space can be written as $ D=(0,\infty)\times R\times R$. It is easy to verify that the MML estimation of $\boldsymbol{\lambda} $ is not at the bound of $D$. So we only consider the consistency on its closed subset $\widetilde{D}=[A_{1},A_{2}]\times[-B,B]\times[-C,C]$, where $A_{1}>0 $ is sufficient small and $A_{2}, B, C  $ are sufficient large.

Let $\widetilde{ Y }=(Y^{\text{obs}},\widetilde{Y}^{\text{mis}})$ and $\widetilde{ Y_{i} } $ be its item response for examinee $i$, for $i=1,\cdots,N$, where $\widetilde{Y}^{\text{mis}}$ is the imputed value in the proposed method. Define 
\begin{equation*}
\begin{split}
&G_{N}(\boldsymbol{\lambda})=\frac{1}{N}\sum_{i=1}^{N}\log P(Y_{i}^{\text{obs}},R_{i}|\boldsymbol{\lambda}),\\
&G(\boldsymbol{\lambda})=\iint\log P(Y ^{\text{obs}},R|\boldsymbol{\lambda})\,dY ^{\text{obs}}\,dR,
\end{split}
\end{equation*}
where $ P(Y ^{\text{obs}},R|\boldsymbol{\lambda})=\iint\sum\limits_{\widetilde{Y}^{\text{mis}}}P(\widetilde{Y},R|\boldsymbol{\lambda},\theta,\tau)P(\widetilde{Y}^{\text{mis}})p(\theta,\tau|\boldsymbol{\mu},\boldsymbol{\Sigma})\,d\theta\,d\tau$.

By Jensen's inequality,
 \begin{equation*}
\begin{split}
\log P(Y ^{\text{obs}},R|\boldsymbol{\lambda})\geqslant &\iint\sum\limits_{\widetilde{Y}^{\text{mis}}}\log \big[P(\widetilde{Y},R|\boldsymbol{\lambda},\theta,\tau)\big]P(\widetilde{Y}^{\text{mis}})p(\theta,\tau|\boldsymbol{\mu},\boldsymbol{\Sigma})\,d\theta\,d\tau\\
=& \sum\limits_{\widetilde{Y}^{\text{mis}}}P(\widetilde{Y}^{\text{mis}})\big[\widetilde{Y}\iint \log F(a(\theta-b))\,d\theta\,d\tau\\
&+(1-\widetilde{Y})\iint \log \big[1-F(a(\theta-b))\big]\,d\theta\,d\tau\big]\\
&+R\iint\log G(c-\tau-C)\,d\theta\,d\tau\\
&+(1-R)\iint\log\big[1-G(c-\tau-C)\big] \,d\theta\,d\tau\\
\end{split}
\end{equation*} 
Using the notations in Assumption~\ref{c1}, therefore,
 \begin{equation*}
\begin{split}
\text{E}\big[\lvert \log P(Y ^{\text{obs}},R|\boldsymbol{\lambda})\rvert\big]\leq&\text{E}\Bigg\{\sum\limits_{\widetilde{Y}^{\text{mis}}}P(\widetilde{Y}^{\text{mis}})\big[\widetilde{Y}\cdot m_{1}(A_{1},B)+(1-\widetilde{Y})\cdot m_{2}(A_{2},B)\big]\\
&+R\cdot m_{3}(C)+(1-R)\cdot m_{4}(C)\Bigg\}\\
=&\big[m_{1}(A_{1},B)-m_{2}(A_{2},B)\big]\text{E}\big[Y^{obs}\big]+m_{2}(A_{2},B)\\
&+\big[m_{3}(C)-m_{4}(C)\big]\text{E}\big[R\big]+m_{4}(C)
\end{split}
\end{equation*} 
Further, by Assumption~\ref{c1},
\[\big[m_{1}(A_{1},B)-m_{2}(A_{2},B)\big]\text{E}\big[Y^{obs}\big]+m_{2}(A_{2},B)\
+\big[m_{3}(C)-m_{4}(C)\big]\text{E}\big[R\big]+m_{4}(C)<\infty  \]
Accordingly, the following uniform law of large numbers holds: 
\[\sup \limits_{\boldsymbol{\lambda}\in \widetilde{D}}\lvert G_{N}(\boldsymbol{\lambda})-G(\boldsymbol{\lambda})\rvert\xrightarrow{P} 0,\]
where $\xrightarrow{P}$ denotes convergence in probability. 
Moreover, by Assumption~\ref{c2} and Gibbs' inequality, $ G(\boldsymbol{\lambda}) $ has a unique maximum at the true parameter $ \boldsymbol{\lambda_{0}} $. Further, by the continuity of $ G(\boldsymbol{\lambda}) $ with respect to $ \boldsymbol{\lambda} $, we
have $ \widehat{\boldsymbol{\lambda}} \xrightarrow{P} \boldsymbol{\lambda_{0}}$.
\begin{flushright}
	$ \hfill\square $
\end{flushright}
\paragraph*{Proof of Theorem~\ref{thm2}}
Since the derivative of $ G_{N}$ with respect to $ (a,b,\zeta) $ at  $ \widehat{\boldsymbol{\lambda}}$ is $ \boldsymbol{0} $ and then by Taylor's expansion, we have that
\renewcommand\arraystretch{1.35}
\begin{equation*}
\sqrt{N}
(\widehat{\boldsymbol{\lambda}}-\boldsymbol{\lambda}_{0})=-\sqrt{N}\boldsymbol{H^{-1}}\big|_{\boldsymbol{\lambda}_{0}}\left(\begin{tabular}{c}
$ \frac{\partial G_{N}}{\partial a}$ \\
$ \frac{\partial G_{N}}{\partial b}$\\
$ \frac{\partial G_{N}}{\partial \zeta}$
\end{tabular}\right)\Bigg|_{\boldsymbol{\lambda}_{0}}+o(1),
\end{equation*}
where $ o(1) $ means the remainder converges to $ \boldsymbol{0} $ in probability and 
\begin{equation*}
 \boldsymbol{H}=\left(\begin{tabular}{ccc}
$ \frac{\partial^{2} G_{N}}{\partial a^{2}}$,&$\frac{\partial^{2} G_{N}}{\partial a \partial b}$,&$\frac{\partial^{2} G_{N}}{\partial a \partial \zeta} $\\
$ \frac{\partial^{2} G_{N}}{\partial a\partial b}$,&$\frac{\partial^{2} G_{N}}{\partial b^{2}}$,&$\frac{\partial^{2} G_{N}}{\partial b \partial \zeta} $\\
$ \frac{\partial^{2} G_{N}}{\partial a \partial\zeta}$,&$\frac{\partial^{2} G_{N}}{\partial b \partial \zeta}$,&$\frac{\partial^{2} G_{N}}{\partial \zeta^{2}} $
\end{tabular}\right).
\end{equation*}
Then, by the multidimensional central limit theorem
, we have
\begin{equation*}
\sqrt{N}\left(\begin{tabular}{c}
$ \frac{\partial G_{N}}{\partial a}$ \\
$ \frac{\partial G_{N}}{\partial b}$\\
$ \frac{\partial G_{N}}{\partial \zeta}$
\end{tabular}\right)\Bigg|_{\boldsymbol{\lambda}_{0}}\xrightarrow{d} \text{N}(\boldsymbol{0},I(\boldsymbol{\lambda_{0}})),
\end{equation*}
where $\xrightarrow{d}$ denote convergence in distribution and 
\begin{equation*}
I(\boldsymbol{\lambda_{0}})=\left(\begin{tabular}{ccc}
$ \text{E}[\frac{\partial p_{\boldsymbol{\lambda}}}{\partial a}]^{2}$,&$\text{E}[\frac{\partial p_{\boldsymbol{\lambda}}}{\partial a}\frac{\partial p_{\boldsymbol{\lambda}}}{\partial b}]$,&$\text{E}[\frac{\partial p_{\boldsymbol{\lambda}}}{\partial a}\frac{\partial p_{\boldsymbol{\lambda}}}{\partial \zeta}]$\\
$ \text{E}[\frac{\partial p_{\boldsymbol{\lambda}}}{\partial a}\frac{\partial p_{\boldsymbol{\lambda}}}{\partial b}]$,&$\text{E}[\frac{\partial p_{\boldsymbol{\lambda}}}{\partial b}]^{2}$,&$\text{E}[\frac{\partial p_{\boldsymbol{\lambda}}}{\partial b}\frac{\partial p_{\boldsymbol{\lambda}}}{\partial \zeta}] $\\
$ \text{E}[\frac{\partial p_{\boldsymbol{\lambda}}}{\partial a}\frac{\partial p_{\boldsymbol{\lambda}}}{\partial \zeta}]$,&$\text{E}[\frac{\partial p_{\boldsymbol{\lambda}}}{\partial b}\frac{\partial p_{\boldsymbol{\lambda}}}{\partial \zeta}]$,&$ \text{E}[\frac{\partial p_{\boldsymbol{\lambda}}}{\partial \zeta}]^{2}$
\end{tabular}\right)\Bigg|_{\boldsymbol{\lambda}_{0}} 
\end{equation*}
is the Fisher information matrix of $ p_{\boldsymbol{\lambda}}=\log P(Y ^{\text{obs}},R|\boldsymbol{\lambda}) $ at $\boldsymbol{\lambda}_{0}$. At the same time, by the weak law of large numbers,
\begin{equation*}
-\boldsymbol{H}\big|_{\boldsymbol{\lambda_{0}}}\xrightarrow{P}-\left(\begin{tabular}{ccc}
$ \text{E}[\frac{\partial^{2} p_{\boldsymbol{\lambda}}}{\partial a^{2}}]$,&$\text{E}[\frac{\partial^{2} p_{\boldsymbol{\lambda}}}{\partial a \partial b}]$,&$\text{E}[\frac{\partial^{2} p_{\boldsymbol{\lambda}}}{\partial a \partial \zeta} ]$\\
$ \text{E}[\frac{\partial^{2} p_{\boldsymbol{\lambda}}}{\partial a\partial b}]$,&$\text{E}[\frac{\partial^{2} p_{\boldsymbol{\lambda}}}{\partial b^{2}}]$,&$\text{E}[\frac{\partial^{2} p_{\boldsymbol{\lambda}}}{\partial b \partial \zeta} ]$\\
$ \text{E}[\frac{\partial^{2} p_{\boldsymbol{\lambda}}}{\partial a \partial\zeta}]$,&$\text{E}[\frac{\partial^{2} p_{\boldsymbol{\lambda}}}{\partial b \partial \zeta}]$,&$\text{E}[\frac{\partial^{2} p_{\boldsymbol{\lambda}}}{\partial \zeta^{2}}] $
\end{tabular}\right)\Bigg|_{\boldsymbol{\lambda}_{0}} =I(\boldsymbol{\lambda_{0}}).
\end{equation*}
Further, by the Continuous Mapping theorem and Slutsky's theorem
 , we have 
\begin{equation*}
\sqrt{N}
(\widehat{\boldsymbol{\lambda}}-\boldsymbol{\lambda}_{0})\xrightarrow{d}\text{N}(\boldsymbol{0},I^{-1}(\boldsymbol{\lambda_{0}})).
\end{equation*}
\begin{flushright}
	$ \hfill\square $
\end{flushright}
\renewcommand\arraystretch{1}
\section*{Appendix B: The detailed sampling procedure}

The detailed sampling procedure using MCMC can be broken down into the following steps:

Step 1: Sample the augmented variable $Z_{ij}$ from the truncated normal
distribution
\[Z_{ij}| \theta_{i},a_{j},b_{j},Y_{ij}\sim
\begin{dcases}
N (a_{j}(\theta_{i}-b_{j}),1)\mathrm{I}(Z_{ij}\geq0)&\text{if $ Y_{ij}=1 $},\\
N (a_{j}(\theta_{i}-b_{j}),1)\mathrm{I}(Z_{ij}<0)&\text{if $Y_{ij}=0$}.
\end{dcases}\]
Step 2: Similarly, sample an additional augmented variable $W_{ij}$ as follow: 
\[W_{ij}| \tau_{i},\zeta_{j},\boldsymbol{\gamma}
,\boldsymbol{R}
_{ij},\boldsymbol{Y}_{ij}\sim
\begin{dcases}
N(\gamma_{0}-\tau_{i}+\zeta_{j}+g(\boldsymbol{R}
_{i,j-1},\gamma_{1})+l(\boldsymbol{Y}_{ij},\gamma
_{2}),1)\mathrm{I}(W_{ij}\geq0) &\text{if $ R_{ij}=1 $},\\
N (\gamma_{0}-\tau_{i}+\zeta_{j}+g(\boldsymbol{R}
_{i,j-1},\gamma_{1})+l(\boldsymbol{Y}_{ij},\gamma
_{2}),1)\mathrm{I}(W_{ij}<0)&\text{if $R_{ij}=0$}.
\end{dcases}\]	
Step 3: Sample the latent ability parameter $\theta_{i}$ for examinee $i$. As previously mentioned, $\left(
\begin{tabular}
[c]{l}%
$\theta_{i}$\\
$\tau_{i}$%
\end{tabular}
\right)  \sim N\left(  \boldsymbol{\mu} ,\boldsymbol{\Sigma}\right)  $. So the prior
distribution of $\theta_{i}$ is the conditional normal distribution%
\[
\theta_{i}\left\vert \tau_{i},\boldsymbol{\mu},\boldsymbol{\Sigma}\right.  \sim N\left(  \mu_{\theta_{i}\left\vert \tau\right.  },\sigma
_{\theta\left\vert \tau\right.  }^{2}\right)  ,
\]
where $\mu_{\theta_{i}\left\vert \tau\right.  }=$ $\mu_{\theta}+\sigma_{\theta
	\tau}\sigma_{\tau}^{-2}\left(  \tau_{i}-\mu_{\tau}\right)  $ and
$\sigma_{\theta\left\vert \tau\right.  }^{2}=\sigma_{\theta}^{2}%
-\sigma_{\theta\tau}^{2}\sigma_{\tau}^{-2}.$ And the full
conditional posterior distribution of $\theta_{i}$ is%
\[
\theta_{i}\left\vert \tau_{i},\boldsymbol{\mu},\boldsymbol{\Sigma}%
,\boldsymbol{Z}_{i\cdot}\right.  \sim N\left(  \mathrm{Var}_{\theta}%
\times\left(  \mu_{\theta_{i}\left\vert \tau\right.  }\sigma_{\theta\left\vert
	\tau\right.  }^{-2}+\sum_{j=1}^{J}a_{j}\left(  Z_{ij}+a_{j}b_{j}\right)
\right)  ,\text{ }\mathrm{Var}_{\theta}\right)  ,
\]
where $\mathrm{Var}_{\theta}=\left(  \sigma_{\theta\left\vert \tau\right.
}^{-2}+\sum_{j=1}^{J}a_{j}^{2}\right)  ^{-1}.$

Step 4: Sample the discrimination parameter $a_{j}$ for item $j$. A prior for
$a_{j}$ is a truncated normal distribution with mean $\mu_{a}$ and variance
$\sigma_{a}^{2}.$ That is, $a_{j}\sim N\left(  \mu_{a},\sigma_{a}^{2}\right)
I\left(  a_{j}>0\right)  .$ Therefore, the full conditional posterior
distribution of $a_{j}$ is%
\[
a_{j}\left\vert \boldsymbol{\theta},b_{j},\boldsymbol{Z}_{\cdot j}\right.  \sim
N\left(  \mathrm{Var}_{a_{j}}\times\left(  \mu_{a}\sigma_{a}^{-2}+\sum
_{i=1}^{N}Z_{ij}\left(  \theta_{i}-b_{j}\right)  \right)  ,\text{
}\mathrm{Var}_{a_{j}}\right)  \mathrm{I}\left(  a_{j}>0\right)  ,
\]
where $\mathrm{Var}_{a_{j}}=\left(  \sigma_{a}^{-2}+\sum\limits_{i=1}%
^{N}\left(  \theta_{i}-b_{j}\right)  ^{2}\right)  ^{-1}$.

Step 5: Sample the difficulty parameter $b_{j}$ for item $j$. A prior for
$b_{j}$ is a normal distribution with mean $\mu_{b}$ and variance $\sigma
_{b}^{2}.$ That is, $b_{j}\sim N\left(  \mu_{b},\sigma_{b}^{2}\right)  .$
Therefore, the full conditional posterior distribution of $b_{j}$ is%
\[
b_{j}\left\vert \boldsymbol{\theta},a_{j},\boldsymbol{Z}_{\cdot j}\right.  \sim
N\left(  \mathrm{Var}_{b_{j}}\times\left(  \mu_{b}\sigma_{b}^{-2}+\sum_{i=1}^{N}a_{j}\left(  a_{j}\theta
_{i}-Z_{ij}\right)  \right)  ,\mathrm{Var}_{b_{j}}\right)  ,
\]
where $\mathrm{Var}_{b_{j}}= \left(  \sigma_{b}^{-2}+Na_{j}^{2}\right)
^{-1}$.

Step 6: Sample the missing response $Y_{ij}^{\text{mis}}$ from
Bernoulli$\left(  q_{ij}\right)$, where $q_{ij}$ was defined in Equation \eqref{qij}.

Step 7: Sample the person missing parameter $\tau_{i}$. Similar to $\theta_{i}$, the conditional prior
distribution of $\tau_{i}$ is 
\[
\tau_{i}\left\vert \theta_{i},\boldsymbol{\mu},\boldsymbol{\Sigma}
\right.  \sim N\left(  \mu_{\tau_{i}\left\vert \theta\right.  },\sigma
_{\tau\left\vert \theta\right.  }^{2}\right)  ,
\]
where $\mu_{\tau_{i}\left\vert \theta\right.  }=$ $\mu_{\tau}+\sigma_{\tau\theta
}\sigma_{\theta}^{-2}\left(  \theta_{i}-\mu_{\theta}\right)  ,$ and
$\sigma_{\tau\left\vert \theta\right.  }^{2}=\sigma_{\tau}^{2}-\sigma_{\theta\tau}^{2}\sigma_{\theta}^{-2}.$ The full conditional
posterior distribution of $\tau_{i}$ is a normal distribution
\[
N\left(\mathrm{Var}_{\tau}\times\left(  \mu_{\tau_{i}|\theta}\sigma_{\tau|\theta }^{-2}+\sum_{j=1}^{J}(  \gamma_{0}+\zeta_{j}+g(\boldsymbol{R}
_{i,j-1},\gamma_{1}) +l(  \boldsymbol{Y}_{ij},\gamma
_{2})-W_{ij} )  \right),\mathrm{Var}_{\tau}\right),
\]
where $\mathrm{Var}_{\tau}=\left(  \sigma_{\tau\left\vert \theta\right.
}^{-2}+J\right)^{-1}$.

Step 8: Sample the item missing parameter $\zeta_{j}$ in the missing mechanism
model. A prior for $\zeta_{j}$ is a normal distribution 
$N\left(  \mu_{\zeta},\sigma_{\zeta}^{2}\right)  .$
Therefore, the full conditional posterior distribution of $\zeta_{j}$ is a normal
distribution %
\[
N\left(\mathrm{Var}_{\zeta}\times\left(  \mu_{\zeta}%
\sigma_{\zeta}^{-2}+\sum_{i=1}^{N}(  -\gamma_{0}+\tau_{i}-g(
\boldsymbol{R}_{ij-1},\gamma_{1})  -l(  \boldsymbol{Y}_{ij},\gamma_{2})  +W_{ij})  \right),\mathrm{Var}_{\zeta}\right),
\]
where $\mathrm{Var}_{\zeta}=\left(  \sigma_{\zeta}^{-2}+N\right)
^{-1}.$

Step 9: Sample the intercept parameter $\gamma_{0}$ in the missing mechanism model. A prior for $\gamma_{0}$ is a truncated normal distribution with mean $\mu_{\gamma_{0}}$ and variance $\sigma_{\gamma_{0}}^{2}.$ That is,
$\gamma_{0}\sim N\left(  \mu_{\gamma_{0}},\sigma_{\gamma_{0}}^{2}\right)
\mathrm{I}\left(  \gamma_{0}<0\right)$. Let
\begin{equation*}
\boldsymbol{E}=
\left(
\begin{tabular}
[c]{l}
$W_{11}+\tau_{1}-\zeta_{1}-g\left(  \boldsymbol{R}_{10},\gamma_{1}\right)  -l\left(
\boldsymbol{Y}_{11},\gamma_{2}\right)  $\\
$W_{12}+\tau_{1}-\zeta_{2}-g\left(  \boldsymbol{R}_{11},\gamma_{1}\right)  -l\left(
\boldsymbol{Y}_{12},\gamma_{2}\right)  $\\
\multicolumn{1}{c}{$\vdots$}\\
$W_{NJ}+\tau_{N}-\zeta_{J}-g\left(  \boldsymbol{R}_{N,J-1},\gamma_{1}\right)  -l\left(
\boldsymbol{Y}_{NJ},\gamma_{2}\right)  $%
\end{tabular}
\right)
  \text{and } 
  \boldsymbol{e}=\left(
\begin{tabular}
[c]{c}%
$e_{11}$\\
$e_{12}$\\
$\vdots$\\
$e_{NJ}$%
\end{tabular}
\right) .
\end{equation*}
where $g\left(\boldsymbol{R}_{i0},\gamma_{1}\right)  =0$. Therefore, 
\[
\boldsymbol{E}=\gamma_{0}\boldsymbol{1}+\boldsymbol{e},
\]
where $\boldsymbol{1}$ is a $N\times J$ dimension vector with elements 1.
So the full conditional posterior distribution of $\gamma_{0}$ is
\[
\gamma_{0}\left\vert \boldsymbol{\tau},\boldsymbol{\zeta},\boldsymbol{W}%
,\boldsymbol{R},\boldsymbol{Y}\right.  \sim N\left(  \mathrm{Var}_{\gamma_{0}}\times\left(  \mu_{\gamma_{0}}\sigma_{\gamma_{0}}^{-2}+\boldsymbol{1}'\boldsymbol{E}\right)  ,\mathrm{Var}_{\gamma_{0}}\right)  \mathrm{I}\left(  \gamma_{0}<0\right) ,
\]
where $\mathrm{Var}_{\gamma_{0}}=\left(  \sigma
_{\gamma_{0}}^{-2}+\boldsymbol{1}'\boldsymbol{1}\right)
^{-1}.$

Step 10: Sample $\gamma_{1}$ in the missing mechanism model. Similar to $\gamma_{0}$, the prior for
$\gamma_{1}$ is $N\left(
\mu_{\gamma_{1}},\sigma_{\gamma_{1}}^{2}\right)  \mathrm{I}\left(  \gamma
_{1}>0\right) .$ Let
\begin{equation*}
\boldsymbol{H}=\left(
\begin{tabular}
[c]{l}
$W_{11}-\gamma_{0}+\tau_{1}-\zeta_{1}-l(  \boldsymbol{Y}_{11},\gamma_{2})  $\\
$W_{12}-\gamma_{0}+\tau_{1}-\zeta_{2}-l(  \boldsymbol{Y}_{12},\gamma_{2})  $\\
\multicolumn{1}{c}{$\vdots$}\\
$W_{NJ}-\gamma_{0}+\tau_{1}-\zeta_{J}-l(  \boldsymbol{Y}_{NJ},\gamma_{2})  $
\end{tabular}
\right)\text{and }\widetilde{\boldsymbol{R}}=\left(
\begin{tabular}
[c]{c}%
$0$\\
$R_{11}$\\
$\vdots$\\
$\sum\limits_{h=1}^{J-1}R_{Nh}$
\end{tabular}
\right). 
\end{equation*}
Accordingly,
\[
\boldsymbol{H}=\gamma_{1}\widetilde{\boldsymbol{R}}+\boldsymbol{e}.%
\]
Therefore, the full conditional posterior distribution of $\gamma_{1}$ is
\[
\gamma_{1}\left\vert \boldsymbol{\tau},\boldsymbol{\zeta},\boldsymbol{W}%
,\boldsymbol{R},\boldsymbol{Y}\right.  \sim N\left(  \mathrm{Var}_{\gamma_{1}}\times\left(  \mu_{\gamma_{1}}\sigma_{\gamma_{1}}^{-2}+\widetilde{\boldsymbol{R}}'\boldsymbol{H}\right)  ,\mathrm{Var}_{\gamma_{1}}\right)  \mathrm{I}\left(  \gamma_{1}>0\right) ,
\]
where $\mathrm{Var}_{\gamma_{1}}=\left(  \sigma
_{\gamma_{1}}^{-2}+\widetilde{\boldsymbol{R}}'\widetilde{\boldsymbol{R}}\right)
^{-1}.$

Step 11: Sample $\gamma_{2}$ in the missing mechanism model. Similar to $\gamma_{0}$ and $\gamma_{1}$, the prior for
$\gamma_{2}$ is  $ N\left(
\mu_{\gamma_{2}},\sigma_{\gamma_{2}}^{2}\right)  \mathrm{I}\left(  \gamma
_{2}<0\right).$ Let
\begin{equation*}
\boldsymbol{K}=\left(
\begin{tabular}
[c]{l}%
$W_{11}-\gamma_{0}+\tau_{1}-\zeta_{1}-g\left(  \boldsymbol{R}_{10},\gamma_{1}\right)
$\\
$W_{12}-\gamma_{0}+\tau_{1}-\zeta_{2}-g\left(  \boldsymbol{R}_{11},\gamma_{1}\right)
$\\
\multicolumn{1}{c}{$\vdots$}\\
$W_{NJ}-\gamma_{0}+\tau_{1}-\zeta_{J}-g\left(  \boldsymbol{R}_{N,J-1},\gamma_{1}\right)  $%
\end{tabular}
\right) \text{and } \widetilde{\boldsymbol{Y}}=\left(
\begin{tabular}
[c]{c}%
$Y_{11}$\\
$Y_{12}$\\
$\vdots$\\
$Y_{NJ}$%
\end{tabular}
\right) .
\end{equation*}
 Equivalently,
\[
\boldsymbol{K}=\gamma_{2}\widetilde{\boldsymbol{Y}}+\boldsymbol{e}.%
\]
Therefore, the full conditional posterior distribution of $\gamma_{2}$ is
\[
\gamma_{2}\left\vert \boldsymbol{\tau},\boldsymbol{\zeta},\boldsymbol{W}%
,\boldsymbol{R},\boldsymbol{Y}\right.  \sim N\left(  \mathrm{Var}_{\gamma_{2}}\times\left(  \mu_{\gamma_{2}}\sigma_{\gamma_{2}}^{-2}+\widetilde{\boldsymbol{Y}}'\boldsymbol{K}\right)  ,\mathrm{Var}_{\gamma_{2}}\right)  \mathrm{I}\left(  \gamma_{2}<0\right) ,
\]
where $\mathrm{Var}_{\gamma_{2}}= \left(  \sigma
_{\gamma_{2}}^{-2}+\widetilde{\boldsymbol{Y}}'\widetilde{\boldsymbol{Y}}\right)
^{-1}.$

Step 12: Sample the covariance $\sigma_{\theta\tau}$ between
$\theta$ and $\tau$. We will use random-walk Metropolis sampling with a truncated normal proposal. At $ m $th iteration, the proposal distribution is $N\left(  \sigma_{\theta\tau}^{\left(
	m-1\right)  },s_{01}^{2}\right)  \mathrm{I}\left(  0<\sigma_{\theta\tau}^{\ast
}<p_{01}\right)  $, where $p_{01}=\sqrt{\sigma_{\tau}^{2,\left(  m-1\right)
}}$ and $s_{01}^{2}$ is the proposal variance. Therefore, the probability of acceptance $\alpha\left(  \sigma
_{\theta\tau}^{\left(  m-1\right)  },\sigma_{\theta\tau}^{\ast}\right)  $ can
be written as%
\[
\min\left\{  1,\frac{\prod\limits_{i=1}^{N}p\left(  \tau_{i}\left\vert
	\theta_{i}^{^{\left(  m\right)  }},\sigma_{\tau}^{2,\left(  m-1\right)
	},\sigma_{\theta\tau}^{\ast}\right.  \right)  p\left(  \sigma_{\theta\tau
	}^{\ast}\right)  \left(  \Phi\left(  \frac{p_{01}-\sigma_{\theta\tau}^{\left(
			m-1\right)  }}{s_{01}}\right)  -\Phi\left(  \frac{-\sigma_{\theta\tau
		}^{\left(  m-1\right)  }}{s_{01}}\right)  \right)  }{\prod\limits_{i=1}%
	^{N}p\left(  \tau_{i}\left\vert \theta_{i}^{^{\left(  m\right)  }}%
	,\sigma_{\tau}^{2,\left(  m-1\right)  },\sigma_{\theta\tau}^{\left(
		m-1\right)  }\right.  \right)  p\left(  \sigma_{\theta\tau}^{\left(
		m-1\right)  }\right)  \left(  \Phi\left(  \frac{p_{01}-\sigma_{\theta\tau
		}^{\ast}}{s_{01}}\right)  -\Phi\left(  \frac{-\sigma_{\theta\tau}^{\ast}%
	}{s_{01}}\right)  \right)  }\right\}  ,
\]
where $p\left(  \tau_{i}\left\vert \theta_{i}\right.  \right)  $ is
the conditional density function of the person parameter in missing model,
and $p\left( \boldsymbol{\cdot}\right) $ is a density of uniform prior.

Step 13: Sample $\sigma_{\tau}^{2}$. Similar to $\sigma_{\theta\tau}$, the random walk Metropolis sampler is also applied to $\sigma_{\tau}^{2}$. The proposal distribution is the truncated normal distribution $N\left(
\sigma_{\tau}^{2,\left(  m-1\right)  },s_{02}^{2}\right)  \mathrm{I}\left(
\sigma_{\tau}^{2,\ast}>p_{0}\right)  $, where $p_{0}=\left(  \sigma_{\theta\tau}^{(m)}\right)  ^{2}$. Therefore, the probability of acceptance $\alpha\left(
\sigma_{\tau}^{2,\left(  m-1\right)  },\sigma_{\tau}^{2,\ast}\right)  $ can be
written as%
\[
\min\left\{  1,\frac{\prod\limits_{i=1}^{N}p\left(  \tau_{i}\left\vert
	\theta_{i}^{^{\left(  m\right)  }},\sigma_{\tau}^{2,\ast},\sigma_{\theta\tau
	}^{\left(  m\right)  }\right.  \right)  p\left(  \sigma_{\tau}^{2,\ast}%
	;\frac{n}{2},\frac{n\alpha}{2} \right)  \left(  1-\Phi\left(  \frac{p_{0}-\sigma_{\tau
		}^{2,\left(  m-1\right)  }}{s_{02}}\right)  \right)  }{\prod\limits_{i=1}%
	^{N}p\left(  \tau_{i}\left\vert \theta_{i}^{^{\left(  m\right)  }}%
	,\sigma_{\tau}^{2,\left(  m-1\right)  },\sigma_{\theta\tau}^{\left(  m\right)
	}\right.  \right)  p\left(  \sigma_{\tau}^{2,\left(  m-1\right)  }%
	;\frac{n}{2},\frac{n\alpha}{2} \right)  \left(  1-\Phi\left(  \frac{p_{0}-\sigma_{\tau
		}^{2,\ast}}{s_{02}}\right)  \right)  }\right\}  ,
\]
where $s_{02}^{2}$ is the proposal variance, and $p\left( \boldsymbol{\cdot};\frac{n}{2},\frac{n\alpha}{2}\right)  $ is the density function of the scaled inverse Gamma distribution IG($ \frac{n}{2} $,$ \frac{n\alpha}{2} $).

What's more, the priors of parameters in above sampling procedure are set as follows:
\[
a_{j}\sim\mathrm{N}\left(  0,1\right)  \mathrm{I}\left(  a_{j}>0\right)
,b_{j}\sim\mathrm{N}\left(  0,1\right)  ,\text{ }\zeta_{j}\sim
\mathrm{N}\left(  0,1\right)  \text{, }j=1,...J.
\]%
\[
\gamma_{0}\sim\mathrm{N}\left(  0,1\right)  \mathrm{I}\left(  \gamma
_{0}<0\right)  \text{, }\gamma_{1}\sim\mathrm{N}\left(  0,1\right)
\mathrm{I}\left(  \gamma_{1}>0\right)  \text{, }\gamma_{2}\sim\mathrm{N}%
\left(  0,1\right)  \mathrm{I}\left(  \gamma_{2}<0\right)
\]%
\[
\mu_{\theta}=\mu_{\tau}=0, s_{01}^{2}=s_{02}^{2}=0.01,\sigma_{\theta\tau}\sim\mathrm{Uniform}\left( 0,1\right)  ,\sigma_{\tau
}^{2}\sim\mathrm{IG}\left(  \frac{0.0001}{2},\frac{0.0001}{2}\right)
\]

\section*{Appendix C: The computational processes of DIC and LPML}

Let $\boldsymbol{\Psi}=\left( \boldsymbol{\tau},\boldsymbol{\zeta},\boldsymbol{\gamma}\right)  $ denote the vector of missing data model parameter in Equation \eqref{pir}. 
Let $\boldsymbol{\Psi}_{ij}=\left( \tau_{i},\zeta_{j},\gamma_{0},\gamma_{1},\gamma_{2}\right)  ^{^{\prime}}$ and $\boldsymbol{\Psi}_{ij}^{\left(  m\right)  }$ be its MCMC sampler from the posterior
distribution at time $m$ for $m=1,...,M$. 

The log-likelihood function of  $ \boldsymbol{R}|\boldsymbol{y} $ evaluated at $\boldsymbol{\Psi
}^{\left(  m\right)}$ is given by
\begin{align*}
\log f\left(\boldsymbol{R}= \boldsymbol{r}\left\vert \boldsymbol{\Psi
}^{\left(  m\right)  }\right. ,\boldsymbol{Y} \right) 
=&\sum\limits_{i=1}^{N}\sum
\limits_{j=1}^{J}\log f\left( r_{ij}\left\vert \boldsymbol{\Psi}_{ij}^{\left(  m\right)  }\right. , Y_{ij} \right)\\
=&\sum\limits_{i=1}^{N}\sum
\limits_{j=1}^{J}[r_{ij}\log (\pi_{ij}^{(m)}) +(1-r_{ij})\log (1-\pi_{ij}^{(m)})]
\end{align*}
where $\pi_{ij}^{(m)}=P\left(  R_{ij}=1
\left\vert \boldsymbol{\Psi}_{ij}^{(m)},\boldsymbol{R}_{ij-1},Y_{ij}\right.
\right) $ and is modeled in Equation \eqref{pir}.

And then the DIC can be
calculated as follows:%
\begin{equation}\label{dic}
\mathrm{DIC}_{\left( \boldsymbol{R}|
	\boldsymbol{Y}  \right)  }=D%
(\boldsymbol{\hat{\Psi}})+2P_{D}={D}(\boldsymbol{\hat{\Psi}}%
)+2\left[  \overline{D}(\boldsymbol{\Psi})-D(\boldsymbol{\hat{\Psi}})\right]  ,
\end{equation}
where 
$\overline{D}(\boldsymbol{\Psi})=-\frac{2}{M}\sum\limits_{m=1}%
^{M}\log f\left(  \boldsymbol{R}\left\vert \boldsymbol{\Psi
}^{\left(  m\right)  }\right.,\boldsymbol{Y}\right)$ and is the EAP of the deviance function $\mathrm{D}%
(\boldsymbol{\Psi})=-2\log f\left(  \boldsymbol{R}\left\vert
\boldsymbol{\Psi}\right.  \right)$. And
$D(\boldsymbol{\hat{\Psi}})=-2\max\limits_{1\leq m\leq M}\log f\left( 
\boldsymbol{R}\left\vert \boldsymbol{\Psi}^{\left(  m\right)  }\right.,\boldsymbol{Y}
\right) 
$ which is an approximation of $\mathrm{Dev}(\boldsymbol{\overline
	{\Psi}})$, when
the prior is relatively non-informative and $\boldsymbol{\overline{\Psi}}$ is the posterior mode. What's more, $P_{D}=\overline{D}(\boldsymbol{\Psi})-D(\boldsymbol{\hat{\Psi}})$ is the
effective number of parameters. Based on the construction, the $\mathrm{DIC}_{\left(  \boldsymbol{R}\left\vert
	\boldsymbol{\Omega}\right.  \right)  }$ given in Equation \eqref{dic} avoids the
situation that $P_{D}$ in traditional definition
may be negative \citep{spiegelhalter2014deviance}.

To calculate the conditional LPML \citep{hanson2011predictive}, the conditional predictive ordinates (CPO; \citealp{hanson2011predictive}) index is needed. A Monte Carlo estimate of the CPO
is given by
\begin{equation}\label{cpo}
\log\widehat{(\mathrm{CPO}_{ij})}=-U_{ij,\max}-\log\left[  \frac{1}{M}
\sum\limits_{m=1}^{M}\exp\left\{  -\log f\left( R_{ij}\left\vert
\boldsymbol{\Psi}_{ij}^{\left(  m\right)  }\right.  ,Y_{ij} \right)  -U_{ij,\max}\right\}
\right]  ,
\end{equation}
where $U_{ij,\max}=\underset{1\leq m\leq M}{\max}\left\{  -\log f\left(
R_{ij}\left\vert \boldsymbol{\Psi}_{ij}^{\left(  m\right)  }\right. ,Y_{ij} \right)
\right\} $. Note that $ U_{ij,\max} $ plays an important role in numerical stabilization in computing
Equation \eqref{cpo}. The LPML is the summary
of $\log(\widehat{\mathrm{CPO}_{ij}})$, so
\[
\mathrm{LPML}_{\left(\boldsymbol{R}\left\vert
	\boldsymbol{\Psi}\right.\right)  }\mathrm{=}\sum\limits_{i=1}^{N}%
\sum\limits_{j=1}^{J}\log\widehat{(\mathrm{CPO}_{ij})}.
\]

\bibliographystyle{apalike}
\bibliography{missing}{}
\end{document}